\documentclass[
reprint,
superscriptaddress,
amsmath,
amssymb,
]{revtex4-2}

\usepackage{graphicx}
\usepackage{dcolumn}
\usepackage{bm}
\usepackage{xcolor}
\usepackage{afterpage}
\usepackage{multirow}
\usepackage{array}
\usepackage{braket}
\usepackage[normalem]{ulem}
\usepackage{hyperref}

\begin{document}

\preprint{APS/123-QED}

\title{Two quantum spin liquid states coexisting in the kagome antiferromagnet herbertsmithite}

\author{Arthur C. Campello}
\affiliation{Stanford Institute for Materials and Energy Sciences, SLAC National Accelerator Laboratory, Menlo Park, CA 94025, USA}
\affiliation{Department of Applied Physics, Stanford University, Stanford, CA 94305, USA}
\author{Shengtao Jiang}
\affiliation{Stanford Institute for Materials and Energy Sciences, SLAC National Accelerator Laboratory, Menlo Park, CA 94025, USA}
\author{Wei He}
\affiliation{Stanford Institute for Materials and Energy Sciences, SLAC National Accelerator Laboratory, Menlo Park, CA 94025, USA}
\author{Jiajia Wen}
\affiliation{Stanford Institute for Materials and Energy Sciences, SLAC National Accelerator Laboratory, Menlo Park, CA 94025, USA}
\author{Daniel M. Pajerowski}
\affiliation{Neutron Scattering Division, Oak Ridge National Laboratory, Oak Ridge, TN 37830, USA}
\author{Steven A. Kivelson}
\affiliation{Department of Physics, Stanford University, Stanford, CA 94305, USA} 
\author{Hong-Chen Jiang}
\affiliation{Stanford Institute for Materials and Energy Sciences, SLAC National Accelerator Laboratory, Menlo Park, CA 94025, USA}
\author{Young S. Lee}
\email{youngsl@stanford.edu}
\affiliation{Stanford Institute for Materials and Energy Sciences, SLAC National Accelerator Laboratory, Menlo Park, CA 94025, USA}
\affiliation{Department of Applied Physics, Stanford University, Stanford, CA 94305, USA} 

\date{\today}

\begin{abstract}
Herbertsmithite (ZnCu$_3$(OH)$_6$Cl$_2$) is one of the most extensively studied quantum spin liquid (QSL) candidates, yet the precise nature of its ground state remains unresolved. 
Here, we compare the results of high-resolution inelastic neutron scattering on single crystals of herbertsmithite (Zn$_\mathrm{x}$Cu$_\mathrm{4-x}$(OD)$_6$Cl$_2$, $x\simeq 0.85$) with large-scale density-matrix renormalization group (DMRG) and dynamical DMRG simulations to benchmark the spin Hamiltonian and uncover fundamental properties of the ground state. The measured spin correlations determine an energy density for the Heisenberg model that closely matches simulations, an essential verification of the Hamiltonian. In parallel, DMRG simulations reveal two nearly degenerate QSL states. Both preserve translational symmetry, but one retains the lattice point-group symmetries whereas the other develops anisotropic spin correlations that break point-group symmetry. These two QSLs have distinct spin structure factors, and, remarkably, close agreement with the neutron spectra is achieved with an admixture of the two states. Our results establish a quantitative correspondence between neutron scattering and simulations, and suggest that herbertsmithite realizes a regime in which two proximate gapped QSL states coexist.
\end{abstract}

\maketitle


\section*{Introduction}
Magnetic frustration in the $S$=1/2 kagome antiferromagnet can stabilize exotic quantum states. Among these are quantum spin liquid (QSL) states, characterized by long-range entanglement and fractionalized excitations~\cite{Balents2010, Savary2016, Broholm2020, shivajiandme}. A better understanding of QSL states might elucidate the physics of unconventional superconductivity~\cite{Anderson1987, Kivelson1987, Laughlin1988, Imajo2021} or help develop topological quantum computers~\cite{Broholm2020, Kitaev2003}.
A key practical challenge is identifying such a QSL state in real materials, whose spectra carry no sharp signature of symmetry breaking and are further obscured by impurities and perturbations beyond the idealized model Hamiltonian. 

Herbertsmithite (ZnCu$_3$(OH)$_6$Cl$_2$), an $S$=1/2 kagome antiferromagnetic (AFM) material, gained attention as a QSL candidate nearly two decades ago due to the absence of magnetic order at temperatures down to $T$=50~mK~\cite{Helton2007, Olariu2008}. No diffraction superlattice peaks indicative of a valence bond crystal (VBC) state were observed~\cite{Freedman2010}, and inelastic neutron scattering measurements showed possible fractionalized spinon excitations consistent with a QSL ground state~\cite{Han2012}. Nuclear magnetic resonance (NMR) and further neutron measurements point to a $\sim$1~meV energy gap in the QSL state~\cite{Fu2015, Han2016, Wang2021}, although different interpretations of the NMR evidence have been argued~\cite{Khuntia2020}. Despite this debate, herbertsmithite remains a preeminent QSL candidate~\cite{Broholm2020, Norman2016,davis2026}. The related $S$=1/2 kagome AFM material Zn-barlowite (ZnCu$_3$(OH)$_6$FBr) provides further evidence for QSL behavior~\cite{Han2014, Jeschke2015, Feng2017, Pasco2018, Smaha2020, Tustain2020, Wang2022, Yuan2022}. Inelastic neutron~\cite{Breidenbach2025} and x-ray~\cite{Smaha2023} scattering reveal a broad continuum of excitations, and NMR~\cite{Wang2021} and neutron~\cite{Breidenbach2025} measurements find both QSL-like excitations and spin gap behavior. The combined results on both Zn-barlowite and herbertsmithite point to universal behavior, strengthening the evidence for QSL physics.

On the theory side, determining the ground state for the $S$=1/2 AFM kagome Heisenberg model is challenging due to the existence of multiple nearly degenerate phases, where such ``intertwined orders'' are ubiquitous in strongly correlated electrons and frustrated magnets~\cite{Fradkin2015, zheng2017hub, ddmrg-trij1j2, kagome-uiuc}. Here, the various competing states include QSLs, VBCs, and magnetically ordered phases. Consensus has been reached that the kagome AFM Heisenberg model with only nearest-neighbor (NN) exchange $J_1$ has a QSL ground state, although its precise nature---whether gapped or gapless---remains under debate~\cite{Jiang2008, Yan2011, Jiang2012, Depenbrock2012, Iqbal2013, Mei2017, Liao2017, He2017, Gong2015, Lauchli2019, taoxiang2026}. The QSL is stable against the inclusion of moderate second NN exchange $J_2$ and 
Dzyaloshinskii-Moriya (DM) interactions,  both of which are present in herbertsmithite and Zn-barlowite, with their magnitudes estimated to lie within the QSL phase boundary~\cite{Jiang2026}. 

Identifying the specific ground state of herbertsmithite therefore requires a quantitative comparison between experiments and theoretical predictions, where several challenges remain. Because a small percentage of excess Cu$^{2+}$ impurity spins exist between the kagome planes, there is some uncertainty about how to model their contribution to the excitation spectra, to what extent the 2D Heisenberg model can capture the remaining spectra, and whether the intrinsic kagome excitations are clearly gapped. Furthermore, the neutron spectra are diffuse in momentum space across different energies~\cite{Han2012,Breidenbach2025}, whereas the calculated low-energy spectra of corresponding QSL states often show pronounced peaks at the $\Gamma'$ positions~\cite{sachdev2014, ddmrg-donna, Jiang2026}.

In this paper, we establish a quantitative correspondence between experiment and large-scale DMRG and dynamical-DMRG simulations, revealing strong evidence for the coexistence of two QSL phases in herbertsmithite.
Advances on both the experimental and numerical fronts made this discovery possible.
A higher signal in high-resolution inelastic neutron scattering on co-aligned single crystals of herbertsmithite enabled new analyses, including a systematic impurity subtraction, which reveals a $\sim 1$~meV gap in the intrinsic kagome spectra. 
The real-space spin correlations extracted from the kagome spectra are in close agreement with simulations, further confirming the applicability of the kagome Heisenberg model to herbertsmithite. 

Furthermore, DMRG calculations uncover two competing QSLs of the kagome Heisenberg model: a symmetric ``isotropic QSL'' that respects the crystal symmetries and a ``chevron QSL'' with anisotropic spin correlations that spontaneously breaks lattice point-group symmetry. Neither QSL alone satisfactorily accounts for the diffuse neutron spectra, since their calculated spectra exhibit peaks at different momenta.  Instead, an approximately equal mixture of both reproduces the neutron spin structure factor. This quantitative agreement provides compelling evidence for two competing QSL states in herbertsmithite, again highlighting the ubiquitous occurrence of intertwined orders in quantum materials, where spatial inhomogeneities enable the coexistence of nearly degenerate phases.

\section*{Structure Factor Comparison and Two QSL States in the Phase Diagram}

\begin{figure}[t]
\includegraphics[width=0.485\textwidth]{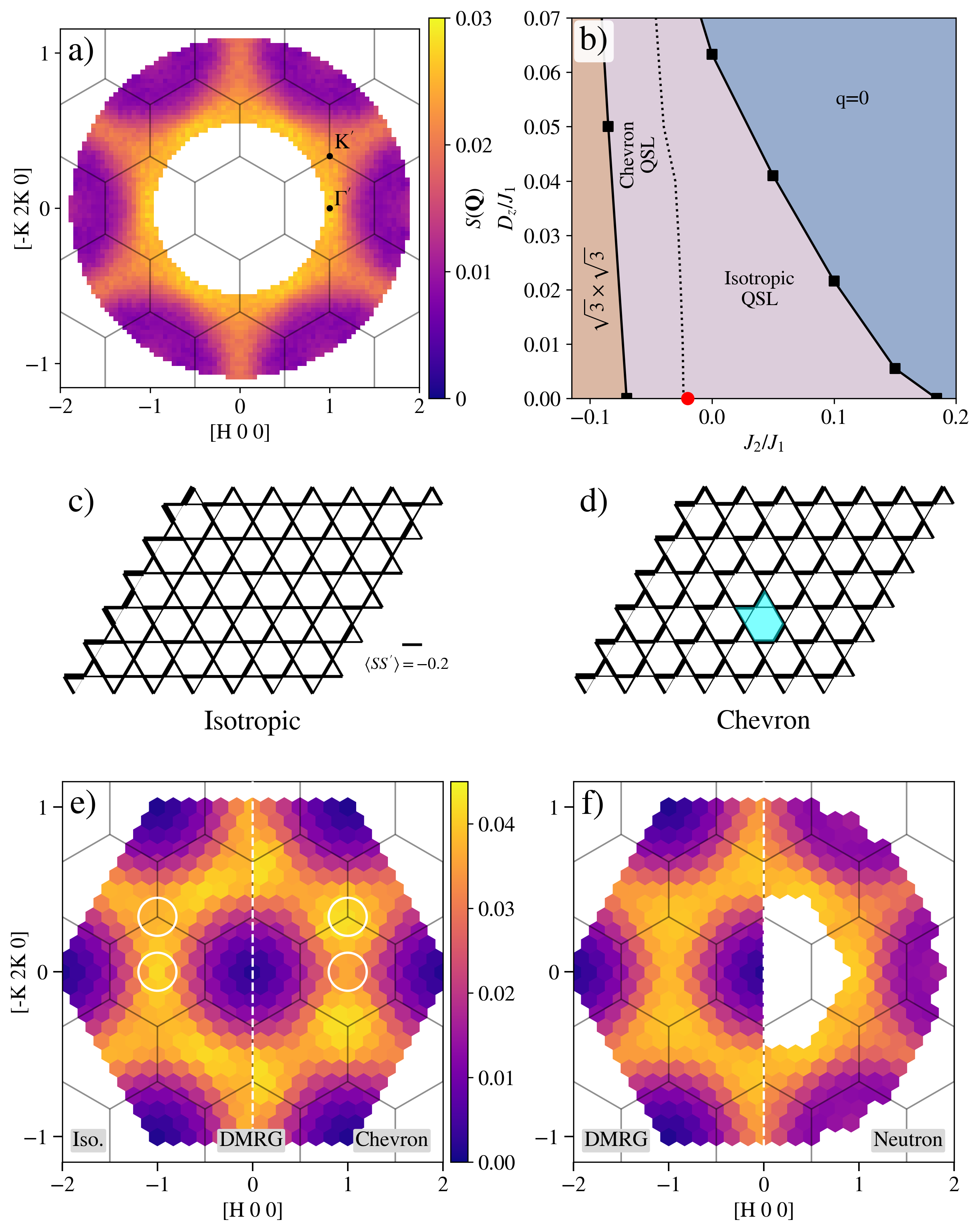}
\caption{\label{fig:two_qsl} \textbf{Isotropic and chevron QSL states.} \textbf{(a)} Neutron scattering intensity integrated from 0 to 8~meV in absolute units per unit cell. \textbf{(b)} Ground state phase diagram of the kagome Heisenberg model in the presence of $J_2$ and $D_z$ couplings, obtained from DMRG simulations on an XC12-12 cylinder. The isotropic and chevron QSLs exist in the intermediate regime. The red point ($J_2=-0.02J_1, D_z=0$) marks the parameters used for comparison with the neutron data in this work. \textbf{(c, d)} Patterns of spin-spin correlation on NN bonds extracted from the bulk of an XC12-12 cluster in DMRG simulations, with their strengths proportional to the thickness of the lines. The isotropic QSL has a more uniform bond pattern, while the chevron QSL has stronger bonds encircling a chevron shape. \textbf{(e)} Equal-time spin structure factor $S(\mathbf{Q})$ for the isotropic (left) and chevron (right) QSL states from DMRG. White circles highlight the differences in $\Gamma'$ and $K'$ intensities between the states. \textbf{(f)} The $S(\mathbf{Q})$ obtained by equally weighting those in (e) (left), compared with the neutron scattering data (right). To facilitate comparison with the neutron data, the $S(\mathbf{Q})$ from DMRG has been $D_6$-symmetrized and multiplied by the magnetic form factor for Cu$^{2+}$.}
\end{figure}

The neutron scattering measurements of this work were performed on a large 2.3~g co-aligned array of high-quality single crystals of deuterated herbertsmithite (Zn$_\mathrm{x}$Cu$_\mathrm{4-x}$(OD)$_\mathrm{6}$Cl$_2$, $x\simeq 0.85$) measured in the (\textit{HK0}) zone. Both neutron energy-loss and energy-gain data were collected using the CNCS spectrometer at the Spallation Neutron Source, which allowed for accurate background subtraction as described in the supplementary material (SM)~\cite{SI}. Panel (a) of Fig.~\ref{fig:two_qsl} shows the resulting data integrated over $\omega=[0,8]$~meV, which approximates the equal-time structure factor $S(\mathbf{Q})$. The scattering intensity reveals a diffuse pattern, consistent with prior neutron measurements of herbertsmithite and with fractionalized excitations~\cite{Han2012}.

The improved signal allows us to make more detailed comparisons of $S(\mathbf{Q})$ with theory. Here, we use DMRG to simulate the spin-$1/2$ kagome Heisenberg model, including second NN exchange $J_2$ and out-of-plane DM interaction $D_z$~\cite{Dzyaloshinsky1958,Moriya1960}, which are expected to be present in these materials~\cite{Han2012,Jeschke2013,Lee2013}. The Hamiltonian is:
\begin{equation}
    \label{Eq:Ham}
H = J_1\sum_{\langle ij\rangle} \mathbf{S}_i\cdot \mathbf{S}_j + J_2\sum_{\langle \langle ij\rangle \rangle} \mathbf{S}_i\cdot \mathbf{S}_j
+ \sum_{\langle ij\rangle}\mathbf{D}_{ij}\cdot (\mathbf{S}_i\times \mathbf{S}_j ).
\end{equation}
The DMRG calculations were performed on an XC12-12 cylinder~\footnote{The XC12-12 cylinder has 12 NN bonds along the circumference and 24 NN bonds along the length.} and kept a maximum bond dimension of around 31000 $U(1)$ equivalent states; further details of the cluster and spin exchanges are provided in the SM~\cite{SI}.

We first map out the ground state phase diagram, as shown in Fig.~\ref{fig:two_qsl}(b). While the phase boundaries to the $q=0$ and $\sqrt3 \times \sqrt 3$~\footnote{For both states, the three spins within one kagome unit cell are oriented at 120$^\circ$ to one another. The $q=0$ state has identical unit cells and therefore preserves translational symmetry. The $\sqrt3 \times \sqrt 3$ state has an extra 120$^\circ$ overall rotation between three unit cells that form a triangle, and therefore has an enlarged unit cell of 9 spins.} magnetically ordered states are consistent with previous studies~\cite{Gong2015, Kolley2015, Jiang2026}, we uncover two {\it distinct} competing QSLs in the intermediate region. These two QSLs differ in their patterns of NN spin-spin correlations, as shown in Figs.~\ref{fig:two_qsl}(c,d). The first QSL has nearly uniform NN spin-spin correlations and preserves all crystal symmetries, so we will refer to it as the ``isotropic QSL.'' The second QSL preserves lattice translations but spontaneously breaks both $C_6$ rotational and inversion symmetries through an anisotropic pattern of NN bond correlations. Its stronger and weaker bonds form the chevron-like motif highlighted in Fig.~\ref{fig:two_qsl}(d); we therefore refer to this state as the ``chevron QSL.'' The transition between these two QSL states appears to be first-order, such that they both remain metastable near the phase boundary. The precise location of the phase boundary is difficult to determine from DMRG for this same reason. Hereafter, we choose $(J_2=-0.02J_1, D_z=0)$ as a representative point in the optimal parameter range relevant to herbertsmithite~\cite{Jiang2026,Breidenbach2025}.

We calculate the equal-time spin structure factor $S(\mathbf{Q})$ for the two QSL phases by Fourier transforming the real-space spin-spin correlation
function $\langle \mathbf{S}_i\cdot\mathbf{S}_j\rangle$.
$S(\mathbf{Q})$ appears similar for both QSL states, with subtle differences: the isotropic QSL exhibits higher intensity at $\Gamma'$, while the chevron QSL has higher intensity at $K'$, as shown in Fig.~\ref{fig:two_qsl}(e). (Calculations without symmetrization appear in the SM~\cite{SI}.) Note that previous studies have also reported or proposed QSLs with $S(\mathbf{Q})$ similar to that of the isotropic QSL, although interpretations of their nature vary~\cite{Kolley2015,Liao2017,sun2024possible,taoxiang2026,dirac-theory,mei2015}. On the other hand, states with $S(\mathbf{Q})$ similar to that of the chevron QSL have also been proposed~\cite{kagome-subir,mei2015}, but are rarely observed in simulations~\cite{kagome-uiuc}. At $(J_2=-0.02J_1, D_z=0)$, these two states are almost degenerate, and their energy densities differ by only $O(10^{-4}J_1)$. On XC12 and XC8 cylinders, both states exhibit finite triplet gaps of approximately $0.11J_1$ and $0.10J_1$, respectively. The weak cylinder-width dependence is consistent with gapped QSLs. 

When compared with the neutron data, both states appear broadly similar, but neither fits perfectly. An equally weighted superposition of the two calculated structure factors removes the peaks and better matches the neutron data, as shown in Fig.~\ref{fig:two_qsl}(f). A quantitative analysis of the residuals appears in the SM~\cite{SI}. This suggests the possible coexistence of these two QSLs in herbertsmithite. As shown later, comparisons with dynamical structure factors in Fig.~\ref{fig:fig4} show greater variations between the two QSL states and provide stronger support for their coexistence.

\section*{Intrinsic Behavior of the Kagome Spin Excitations}

\begin{figure}[b]
\includegraphics[width=0.485\textwidth]{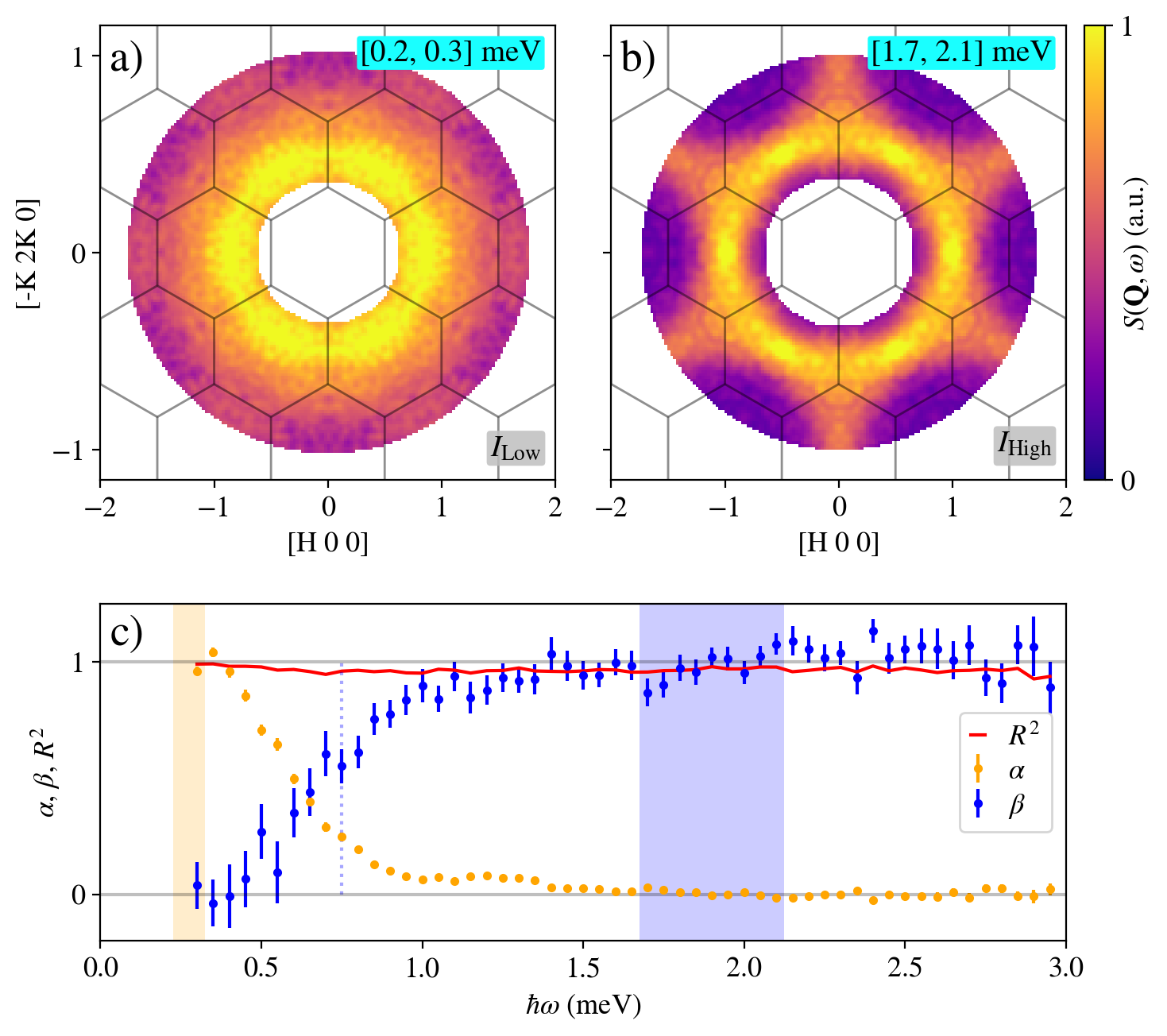}
\caption{\label{fig:fig_two_slice}
\textbf{Separation of impurity and intrinsic kagome scattering.}
\textbf{(a,b)} Scattering contours in the $(\textit{HK0})$ plane at $T=1.7$~K integrated over $0.2\leq\omega\leq0.3$~meV and $1.7\leq\omega\leq2.1$~meV, respectively. Both panels use the same intensity scale. The low- and high-energy slices define the two basis patterns used in the linear decomposition.
\textbf{(c)} Fitted coefficients $\alpha$ and $\beta$, together with the coefficient of determination $R^2$, obtained by modeling each energy slice as $\alpha I_\mathrm{Low}+\beta I_\mathrm{High}$. Shaded regions mark the energy windows used to define the basis slices. The dotted line marks the energy at which $\beta$ is closest to 0.5.}
\end{figure}

The high quality of the inelastic neutron scattering signal allows us to distinguish the intrinsic kagome contribution from the impurity contribution with minimal assumptions. Panels (a) and (b) in Fig.~\ref{fig:fig_two_slice} show the scattering intensity measured at the base temperature $T$=1.7~K for two representative energy slices. The higher-energy spectrum shows a characteristic ``star'' pattern also observed in Zn-barlowite~\cite{Breidenbach2025}, which arises predominantly from NN spin-spin correlations. Lower-energy scattering at the base temperature reveals a pattern known to arise from impurity correlations~\cite{Han2016}.

If these two patterns represent two distinct contributions to the excitation spectra, one should be able to model every energy slice $I(\mathbf{Q},\omega)$ as a simple linear combination of the two: $\alpha(\omega) I_\mathrm{Low}+\beta(\omega) I_\mathrm{High}$. Panel (c) of Fig.~\ref{fig:fig_two_slice} shows the fitted coefficients with the fit quality quantified by $R^2$. These two components describe the data well over the entire energy range (measured with $E_i=3.32$~meV). The impurity-associated coefficient $\alpha$ falls below $10\%$ by $1$~meV and is negligible by $1.5$~meV, while the intrinsic-kagome-associated $\beta$ value exhibits a gap-like onset, rising from near zero below $0.4$~meV to 0.5 near $0.8$~meV and reaching unity by $1.5$~meV. This decomposition shows that, with minimal assumptions, the intrinsic spectral weight exhibits gap-like behavior below $1$~meV.

\begin{figure}[t]
\includegraphics[width=0.485\textwidth]{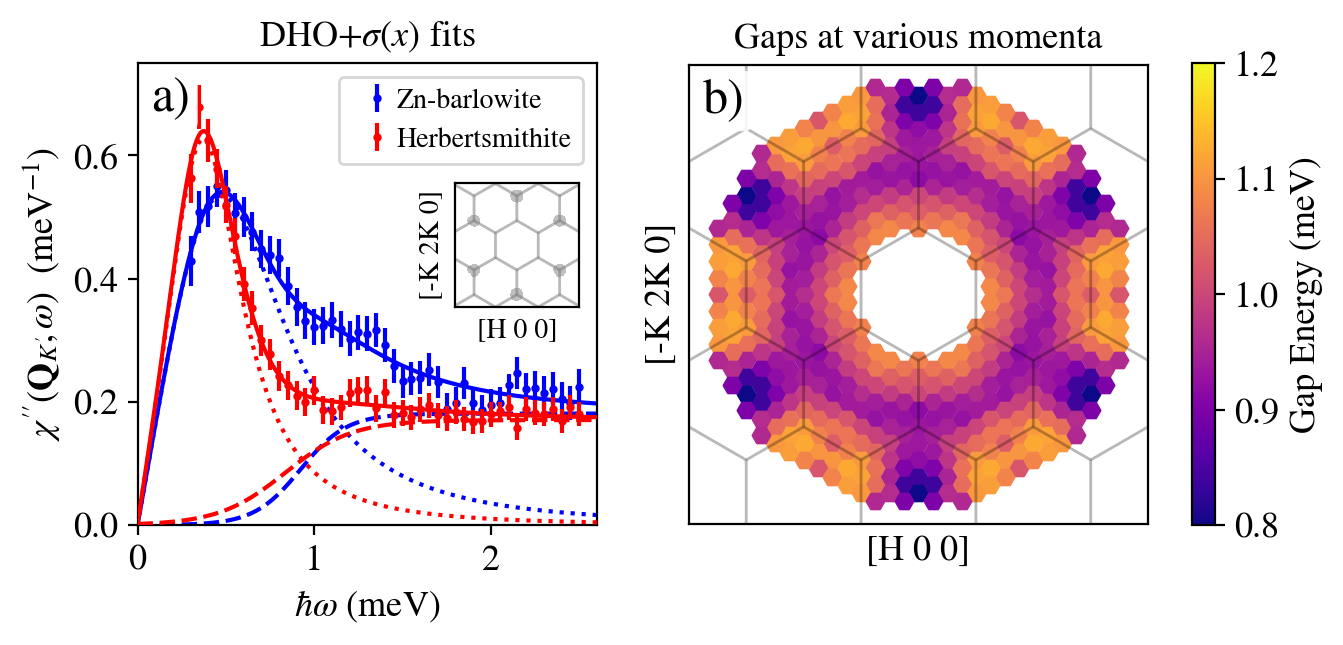}
\caption{\label{fig:fig2} \textbf{Intrinsic spin gap in $S=1/2$ kagome materials, including momentum dependence of the gap.} \textbf{(a)} Energy dependence of $\chi''(\omega)$ at the $K'$ point for herbertsmithite and Zn-barlowite~\cite{Breidenbach2025} in absolute units. DHO+$\sigma(x)$ fits are shown in solid lines with the DHO components in dotted lines and the sigmoid components in dashed lines. The inset shows the momentum region of integration. \textbf{(b)} Spin gap energy extracted from the sigmoid fit mapped at various momenta.}
\end{figure}

We can further quantify the spin gap, including its momentum dependence, by assuming a simple empirical model for the impurity contribution. Here, we convert $S(\mathbf{Q},\omega)$ to the imaginary part of the dynamic susceptibility $\chi''(\mathbf{Q},\omega)$ using the fluctuation-dissipation relation and fit selected energy cuts with two components: a damped-harmonic-oscillator (DHO) response representing the impurity-dominated low-energy mode, and a sigmoid-like onset representing the intrinsic kagome contribution. This phenomenological decomposition follows the approach used in our prior studies of herbertsmithite and Zn-barlowite~\cite{Han2016, Breidenbach2025}, but here the higher statistics allow the gap energy to be tracked across momentum. Panel (a) of Fig.~\ref{fig:fig2} shows $\chi''(\omega)$ near the $K'$ momentum points for herbertsmithite and Zn-barlowite~\cite{Breidenbach2025}. The DHO+$\sigma(x)$ fit reveals that while the overall signals differ significantly due to expected impurity-environment differences, the sigmoid-like intrinsic kagome contributions of the two materials are similar.

In fact, it is quite gratifying to quantitatively observe a clear difference in the low-energy DHO behavior between herbertsmithite and Zn-barlowite, since the different impurity environments lead to different couplings between the impurity and kagome moments. The DHO peak occurs at 0.45(5)~meV in herbertsmithite and 0.65(2)~meV in Zn-barlowite. Additionally, the ratio of the integrated DHO intensities is 0.76, closely matching the ratio of the estimated impurity concentrations of 15\%\;$:20$\%. This further justifies an impurity origin for the low-energy response, whereas the similar sigmoid amplitudes are consistent with an intrinsic contribution from the
kagome planes~\cite{Han2016, Breidenbach2025}.

The higher statistics of the herbertsmithite data allow us to uncover the momentum dependence of the energy gap. Figure~\ref{fig:fig2}(b) shows this dependence derived by constraining a common damping and resonance for the DHO component and letting all other parameters fit freely. The average gap energy is 1.0(1)~meV; see the SM for more details. The gap energy exhibits a measurable momentum dependence which appears to be correlated with the intensity pattern of $S(\mathbf{Q})$ (higher intensity corresponds to smaller gap values). Our simulations also show a  weak momentum dependence of the gap with a minimum at $\Gamma'$, again suggesting that the gap is related to the intrinsic kagome moments.

\section*{Matching Experiment and Theory: Evidence for Two QSL States in the Herbertsmithite Hamiltonian}

\begin{figure}[t]
\includegraphics[width=0.485\textwidth]{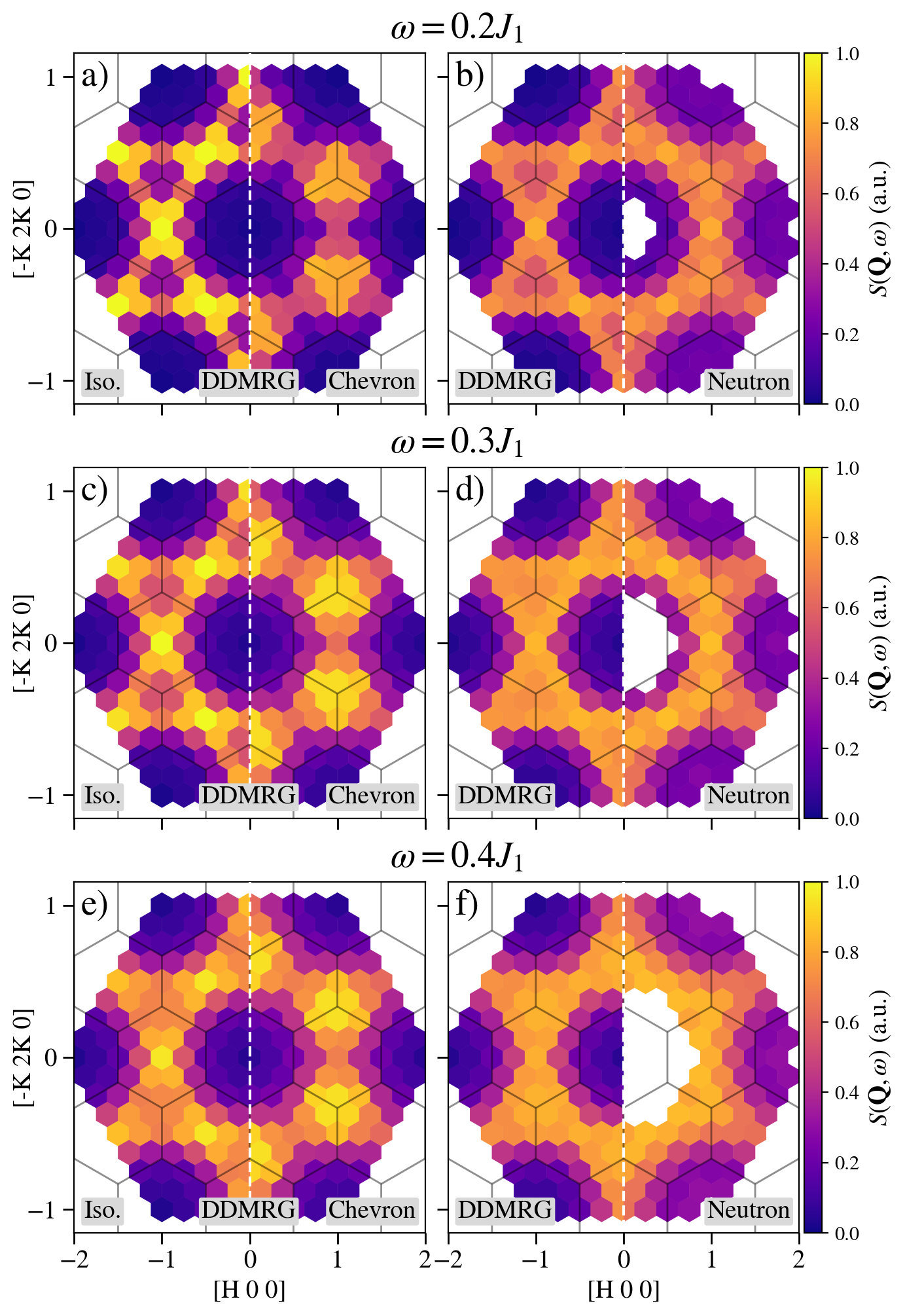}
\caption{\label{fig:fig4} \textbf{Isotropic and chevron QSLs. Comparison of dynamical structure factors from DDMRG and neutron scattering.} \textbf{(a, c, e)} DDMRG $S(\mathbf{Q},\omega)$ corresponding to isotropic (left) and chevron (right) QSL states for $\omega$=0.2, 0.3, 0.4$J_1$, respectively. \textbf{(b, d, f)} Equally mixed, symmetrized isotropic and chevron states calculated using DDMRG (left) compared against neutron scattering data at various energies, $\omega$=0.2, 0.3, 0.4$J_1$. The DDMRG simulations are performed on XC8-12 cylinders with $(J_2=-0.02J_1, D_z=0)$ and a broadening factor $\eta=0.05J_1$, keeping a maximum bond dimension of 2400. Further details regarding the DDMRG algorithm, its convergence, and results on the XC12 cylinder are shown in the SM~\cite{SI}. The DDMRG calculations are scaled by factors of 0.52, 0.28, and 0.24, respectively, to match the neutron data in intensity. Magnetic form factors are applied to all DDMRG plots for comparison to neutron data.}
\end{figure}

We next compare the measured energy-resolved spectra directly with simulations of the dynamical spin structure factor $S(\mathbf{Q},\omega)$ using an improved dynamical-DMRG (DDMRG) method~\cite{ddmrg-jeckelmann, ddmrg-trij1j2}. The results are shown in Fig.~\ref{fig:fig4}(a, c, e) for $\omega$=0.2, 0.3, and 0.4$J_1$, respectively. The two QSL states (isotropic and chevron) exhibit clearly distinguishable features, especially for $\omega=0.2J_1$. The isotropic QSL exhibits higher intensity at the $\Gamma'$ point of the Brillouin zone, while the chevron QSL shows intensity more concentrated around the $K'$ points.  The contrast here for the low-energy $S(\mathbf{Q},\omega)$ is more dramatic than that in the $S(\mathbf{Q})$ shown in Fig.~\ref{fig:two_qsl}.
The $S(\mathbf{Q},\omega)$ also exhibits a spin triplet gap at $0.1$--$0.15J_1$~\cite{SI}, consistent with the $0.11J_1$ gap obtained in DMRG. 

The right halves of panels (b, d, f) of Fig.~\ref{fig:fig4} show the neutron data plotted with the same momentum resolution and matched in scale for the respective energies. The signal is diffuse in momentum space without any dominant peaks. Neither of the two individual DDMRG spectra matches this, especially at low energy $\omega=0.2J_1$, where they exhibit clear peaks at $\Gamma'$ or around $K'$, respectively. However, if we superpose the simulated $S(\mathbf{Q},\omega)$ of these two states with equal weight~\footnote{The equal mixture is very close to the optimal value in terms of matching the neutron data as measured by an $R^2$ shown in the SM~\cite{SI}.}---shown in the left halves of panels (b, d, f)---we see remarkable agreement with the neutron data. 
This close agreement is compelling evidence that both isotropic and chevron-like QSL correlations contribute to the measured dynamical magnetic response. As far as we know, there does not exist a single state that matches the measured $S(\mathbf{Q},\omega)$ that has a diffuse structure from $\Gamma'$ to $K'$, even for the proposed QSL with a spinon Fermi surface~\cite{mei2015}. The two QSLs observed in our simulations appear closest to the two gapped $\mathbb{Z}_2$ QSLs proposed in Ref.~\cite{kagome-subir} and previously reported in an exact diagonalization study~\cite{kagome-uiuc}, although there was no mention of the chevron pattern or bond nematicity.

\begin{figure}[t]
\includegraphics[width=0.485\textwidth]{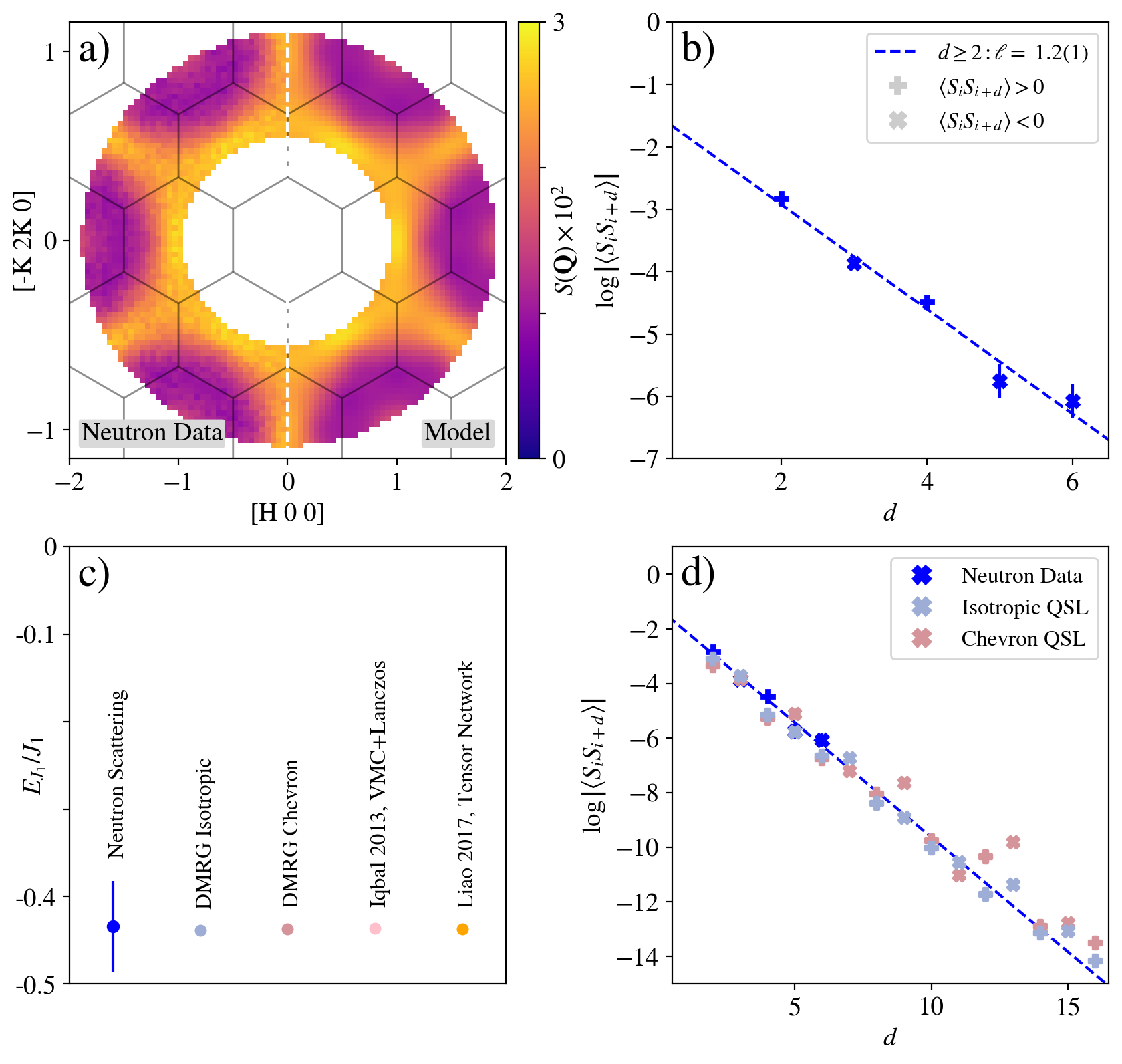}
\caption{\label{fig:spin_corrs} 
\textbf{Spin-spin correlations, derived ground state energy density and fitted correlation length.} \textbf{(a)} Left half: Intensity integrated from 0 to 8~meV. Right half: Intensity from the pair-correlation model including bond correlations up to $d$=6 NN spacings. \textbf{(b)} The absolute spin-spin correlations versus distance in units of NN distance $d$, on a log-linear scale. Symbols show whether correlations are positive or negative. An error-weighted linear fit of $d\geq 2$ correlations implies an exponential correlation length of $\sim$1.2 lattice spacings. \textbf{(c)} Energy density derived from the NN spin-spin correlation from neutron scattering data compared with selected numerical calculations for the NN kagome Heisenberg model. The error bar incorporates counting statistics, regression uncertainty, and uncertainty in the impurity levels. \textbf{(d)} Spin-spin correlations in panel (b) compared to those for both simulated QSL states up to 16 NN spacings.}
\end{figure}

The broad momentum coverage of the neutron measurements also allows the comparison of real-space spin correlations between experiment and theory. We use the intensity integrated from 0 to 8~meV as an estimate of the equal-time structure factor and extract average spin-spin correlations using a pair-correlation model~\cite{Breidenbach2025}. The pair-correlation model expresses the momentum dependence of the form-factor-corrected intensity as a linear combination of correlations
between symmetry-equivalent spin pairs; details of the model and fitting procedure are provided in the SM~\cite{SI, Han2016, Breidenbach2025}. Figure~\ref{fig:spin_corrs}(a) compares the measured structure factor with the pair-correlation reconstruction using correlations extending to $d=6$ NN spacings. The excellent agreement indicates that the dominant correlations are short-ranged and are well described by the model.

Our data allow for the determination of the spin correlation length, even in the absence of well-defined peaks. Figure~\ref{fig:spin_corrs}(b) shows the absolute value of the extracted spin-spin correlations along a lattice direction as a function of distance. We perform a linear fit on a logarithmic scale for $d\geq2$, excluding the NN correlation to reduce the influence of local physics. Errors are propagated assuming independent regression components. An exponential decay of the spin-spin correlations fits the data well and yields a correlation length of $\ell\simeq1.2(1)$ NN spacings. This exponential real-space decay is further evidence for the gapped nature of the ground state. A power-law description of the data does not fit as well~\cite{SI}.

Finally, as a crucial verification, our neutron results directly confirm the appropriateness of the nearest-neighbor Heisenberg model as a description of the herbertsmithite Hamiltonian. The spin-spin correlation $\langle \mathbf{S}_i \cdot \mathbf{S}_j \rangle_{NN}$ is the dominant contribution to the measured signal in Fig.~\ref{fig:spin_corrs}(a) and can be converted into the corresponding energy density $E_{J_1}/J_1$ associated with $J_1$ exchange. The experimentally derived $E_{J_1}/J_1$ is plotted in Fig.~\ref{fig:spin_corrs}(c) and agrees, within error, with values from various numerical calculations, including DMRG, tensor-network~\cite{Liao2017}, and variational Monte Carlo~\cite{Iqbal2013}, as well as other calculations~\cite{Ran2007,Lauchli2019}. This agreement provides an essential check that herbertsmithite is accurately described by the kagome Heisenberg model with a dominant NN spin exchange. In contrast, in barlowite 2, which has additional ferromagnetic interlayer exchange, the experimentally derived $E_{J_1}/J_1$ is only 0.3.\footnote{This value is presented in a forthcoming paper by He et al.} Of course, the measured $E_{J_1}/J_1$ in herbertsmithite cannot distinguish among the various competing states, which are exceedingly close in energy. However, it provides a necessary condition that the neutron spectra of any real material claimed to be described by the spin-$1/2$ kagome Heisenberg model must satisfy.

Figure~\ref{fig:spin_corrs}(d) shows the spin correlation data from panel (b) with the exponential decay fit extended to $d$=16 and the spin correlations from DMRG simulations for both the isotropic and chevron QSLs. Both QSLs show a short correlation length around 1.2 NN spacings, which aligns closely with that estimated from the neutron data and is consistent with a gapped QSL. The agreement in both the magnitude of the NN correlation and the decay length provides additional independent consistency checks for the two-QSL interpretation inferred from the momentum-space structure factors.

\section*{Discussion}
We have shown that high-resolution neutron scattering combined with DMRG and DDMRG establishes a quantitative connection between the magnetic response of herbertsmithite and QSL states in the kagome Heisenberg model. After carefully separating the impurity-dominated low-energy signal, the intrinsic kagome response shows a gap-like onset near 1~meV and diffuse spectra above this.
Two competing QSLs are uncovered in DMRG simulations of the kagome Heisenberg model, with one preserving the point-group symmetry and the other breaking it. Neither alone reproduces the neutron dynamical structure factor, whereas an approximately equal admixture captures both its momentum and energy dependence. 

Nearly degenerate phases are not uncommon in strongly correlated systems, and a similar scenario of competing QSLs has also recently been found in the triangular-lattice Heisenberg model~\cite{ddmrg-trij1j2}. In a real crystal, the local strain around an impurity could give rise to weak spatial variations in exchange couplings, and may therefore locally favor different QSLs and facilitate the coexistence of both in the material. Notably, NMR measurements on Zn-barlowite have revealed evidence of lattice distortion in an appreciable fraction of the sample, which breaks the point-group symmetry of the kagome lattice and can favor the chevron QSL in those regions~\cite{Yuan2022,Wang2022}.

Controlled perturbations that couple to the point-group symmetry breaking of the chevron state can provide direct tests of this interpretation. For example, uniaxial strain should favor two anti-parallel chevron orientations with a specific nematic axis and redistribute spectral weight between the $\Gamma'$- and $K'$-centered responses; the strong strain sensitivity observed in Y-kapellasite demonstrates that small distortions can select among competing kagome states~\cite{Wang2023}. Inelastic x-ray scattering has revealed unusual phonon dynamics in small Zn-barlowite crystals~\cite{Campello2025}, and measurements under strain may prove insightful. Additionally, for the chevron state, in the context of an underlying Hubbard-like model with charge degrees of freedom, the redistribution of spin correlations from weak to strong bonds implies a local ferroelectric moment. Electric-field cooling could therefore provide an additional means of selecting a single domain, although the associated electric dipole moment is small and a large electric field may be required~\footnote{The electric dipole moment should be proportional both to $t/U$ for an underlying Hubbard model with NN hopping $t$ and onsite repulsion $U$ and to the magnitude of the order parameter.}. Additionally, optical measurements, including intriguing results from second harmonic generation~\cite{Laurita2019}, can help identify broken symmetries in herbertsmithite. Such experiments would test whether the near-degeneracy of the two QSLs can be lifted to selectively stabilize one state.

\section*{Acknowledgments}
We thank Takashi Imai, Steve White, Xue-Yang Song,  Hitesh Changlani, Mike Norman, David Hsieh, Rebecca Smaha, Emily Zhang and Emma Rabinowitz for insightful discussions. This work is supported by the U.S. Department of Energy (DOE), Office of Science, Basic Energy Sciences, Materials Sciences and Engineering Division, under contract DE-AC02-76SF00515. This research used resources at the Spallation Neutron Source, Department of Energy (DOE), Office of Science User Facilities, operated by the Oak Ridge National Laboratory.

\section*{Competing Interests}
The authors declare no competing interests.

\section*{Author Contributions}
ACC, SJ, HCJ, and YSL conceived the project. ACC performed the neutron measurements with help from YSL, DMP, and WH. ACC performed the analysis and created the figures with input from SJ and HCJ. SJ and HCJ performed the numerical simulations. WH and JW synthesized and characterized the herbertsmithite single crystals and performed the sample assembly. ACC, SJ, HCJ, SAK, and YSL wrote the paper with input from all co-authors.

\bibliography{refs}

\end{document}


\preprint{APS/123-QED}

\title{Supplementary Material for: Two quantum spin liquid states coexisting in the kagome antiferromagnet herbertsmithite}

\author{Arthur C. Campello}
\affiliation{Stanford Institute for Materials and Energy Sciences, SLAC National Accelerator Laboratory, Menlo Park, CA 94025, USA}
\affiliation{Department of Applied Physics, Stanford University, Stanford, CA 94305, USA}
\author{Shengtao Jiang}
\affiliation{Stanford Institute for Materials and Energy Sciences, SLAC National Accelerator Laboratory, Menlo Park, CA 94025, USA}
\author{Wei He}
\affiliation{Stanford Institute for Materials and Energy Sciences, SLAC National Accelerator Laboratory, Menlo Park, CA 94025, USA}
\author{Jiajia Wen}
\affiliation{Stanford Institute for Materials and Energy Sciences, SLAC National Accelerator Laboratory, Menlo Park, CA 94025, USA}
\author{Daniel M. Pajerowski}
\affiliation{Neutron Scattering Division, Oak Ridge National Laboratory, Oak Ridge, TN 37830, USA}
\author{Steven A. Kivelson}
\affiliation{Department of Physics, Stanford University, Stanford, CA 94305, USA} 
\author{Hong-Chen Jiang}
\affiliation{Stanford Institute for Materials and Energy Sciences, SLAC National Accelerator Laboratory, Menlo Park, CA 94025, USA}
\author{Young S. Lee}
\email{youngsl@stanford.edu}
\affiliation{Stanford Institute for Materials and Energy Sciences, SLAC National Accelerator Laboratory, Menlo Park, CA 94025, USA}
\affiliation{Department of Applied Physics, Stanford University, Stanford, CA 94305, USA} 

\date{\today}

\maketitle

\onecolumngrid
\section{Further Details of Sample and Analysis of Neutron Data}

\subsection{Herbertsmithite sample details}

\begin{figure}[htbp]
\centering
\includegraphics[width=0.35\textwidth]{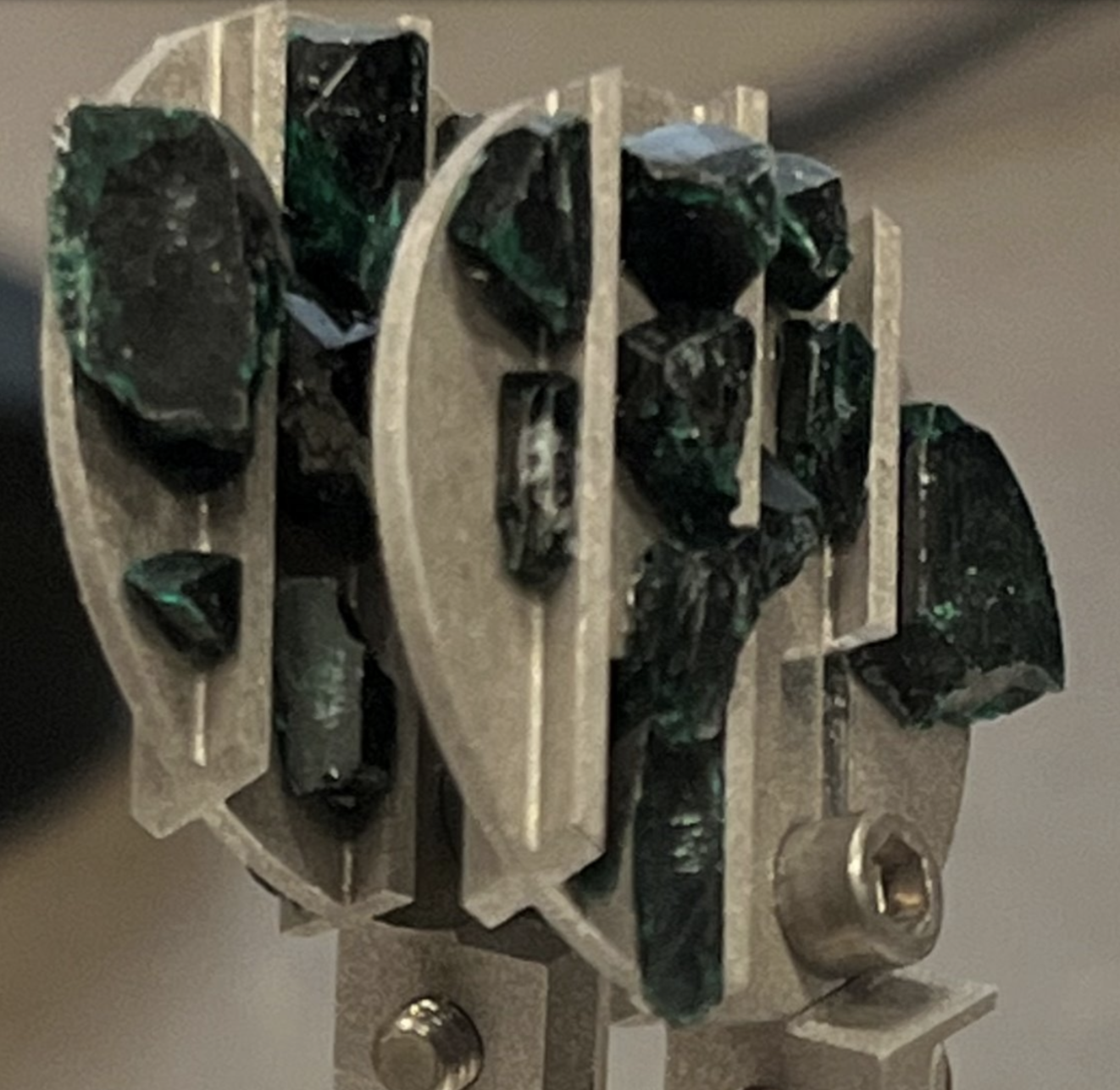}
\caption{\textbf{Co-aligned herbertsmithite crystal array.}
Photograph of the 2.3~g array of 26 deuterated herbertsmithite single
crystals used for the neutron scattering measurements. The crystals
were co-aligned to provide access to the $(HK0)$ scattering plane.}
\label{fig:sample_array}
\end{figure}

The neutron-scattering measurements were performed on a 2.3~g
co-aligned array of 26 deuterated herbertsmithite single crystals,
$\mathrm{Zn}_{x}\mathrm{Cu}_{4-x}(\mathrm{OD})_6\mathrm{Cl}_2$ with
$x \simeq 0.85$, shown in Fig.~\ref{fig:sample_array}. The crystals
were co-aligned to provide access to the $(HK0)$ scattering plane.

To assess sample-to-sample consistency and establish the characteristic
magnetic exchange energy scale, dc magnetic susceptibility was measured for
three representative clusters of crystals (each cluster contains around 20 pieces of randomly oriented small single crystals) in an applied field of $\mu_0 H=5$~T.
After subtraction of the empty-straw background, the data between
180 and 370~K were fit to
\begin{equation}
    \chi(T)=\chi_0+\frac{C}{T-\Theta_{\mathrm{CW}}}.
\end{equation}
The Curie constant $C$ was constrained using $S=1/2$, 3.15 Cu$^{2+}$
ions per formula unit, and $g=2.22$, the orientationally averaged
value reported for single-crystal herbertsmithite
\cite{Han2012a}. The resulting Curie-Weiss temperatures $\Theta_{\mathrm{CW}}$ range
from $-249$ to $-256$~K, with a mean of $-253$~K and a
sample-to-sample standard deviation of approximately 4~K. Their close
agreement indicates a consistent dominant antiferromagnetic exchange
scale among the measured crystals.

The Curie-Weiss temperature is not itself a direct determination of
$J_1$ because the fitting range is comparable to $J_1/k_{\mathrm B}$
and high-temperature corrections are significant. Previous
herbertsmithite measurements found
$\Theta_{\mathrm{CW}}=-300(20)$~K and $J_1 \simeq 17$~meV
\cite{Helton2007}. Using this result as an approximate experimental
calibration, the mean Curie-Weiss temperature of the present crystals
corresponds to $J_1 \simeq 14$--15~meV. This is also consistent with
first-principles estimates of $J_1 \simeq 16$~meV
\cite{Jeschke2013}. We therefore use $J_1=15$~meV as the representative
exchange scale when comparing the experimental and numerical energy
axes.

\begin{table}[htbp]
\centering
\caption{\textbf{Curie-Weiss fits for representative herbertsmithite
crystals.} Fit parameters obtained from dc susceptibility measured in
an applied field of $\mu_0H=5$~T. The data were fit between 180 and
370~K after subtraction of the empty-straw background. The number of
Cu$^{2+}$ ions per formula unit and the $g$ factor were fixed at 3.15
and 2.22, respectively.}
\label{tab:hbs_curie_weiss}
\begin{tabular}{lcccccc}
\toprule
Sample &
Mass (mg) &
Cu$^{2+}$ per f.u. &
$g$ &
$\Theta_{\mathrm{CW}}$ (K) &
$\chi_0$ ($10^{-4}$ emu/mol) &
Reduced $\chi^2$ \\
\midrule
1   & 52.9 & 3.15 & 2.22 & $-256$ & 1.1 & 3.4  \\
2   & 49.7 & 3.15 & 2.22 & $-249$ & 1.4 & 5.2  \\
3 & 50.0 & 3.15 & 2.22 & $-254$ & 1.2 & 10.5 \\
\bottomrule
\end{tabular}
\end{table}

\subsection{Neutron-data reduction and background subtraction}

\begin{figure*}[htbp]
\centering
\includegraphics[width=0.98\textwidth]{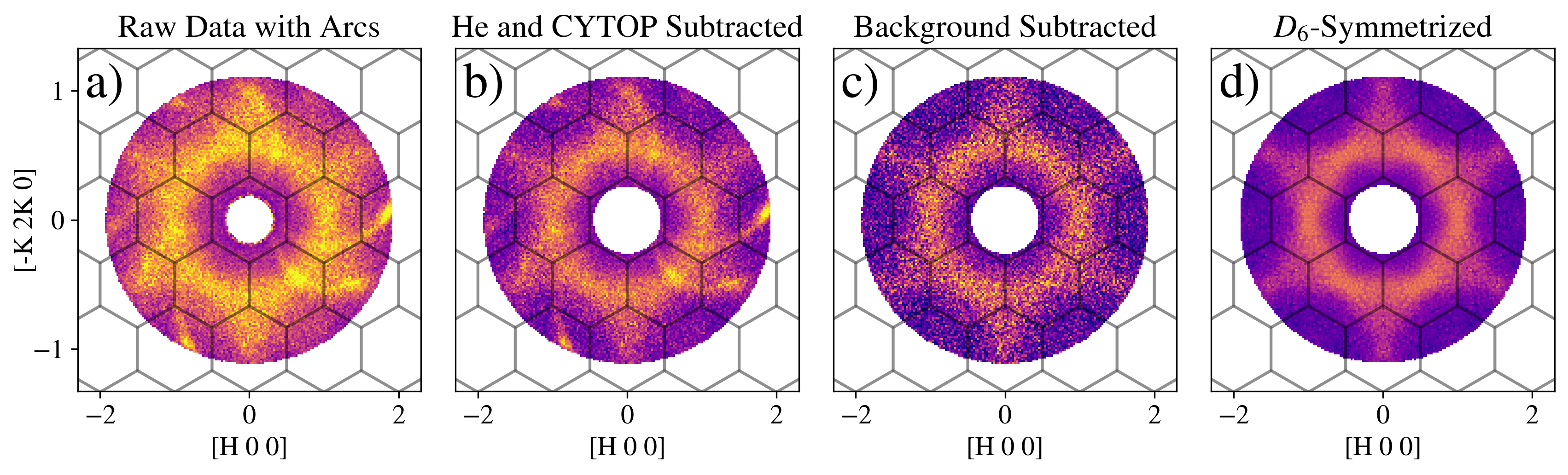}
\caption{\textbf{Neutron-data processing sequence.}
Representative $T=1.7$~K scattering integrated over $1.0 \leq \hbar\omega \leq 1.5$~meV. (a) Raw reduced CNCS data, including arc-like parasitic scattering. (b) Data after subtraction of an empty sample-environment measurement containing the helium and CYTOP contributions. (c) Data after removal of the remaining temperature-independent background using the detailed-balance procedure described in the text.
(d) Final data after $D_6$ symmetrization in the $(HK0)$ plane. Gray lines indicate Brillouin-zone boundaries.}
\label{fig:si_data_processing}
\end{figure*}

The neutron scattering data were collected on the Cold Neutron Chopper Spectrometer (CNCS) at the Spallation Neutron Source. Measurements (SNS) at Oak Ridge National Laboratory with incident energy $E_i=3.32$~meV were performed at $T=1.7$, 40, and 150~K, while measurements with $E_i=12$~meV were performed at $T=1.7$ and 40~K. The $E_i=3.32$~meV measurements used a chopper frequency of 180~Hz. At each temperature, the sample was rotated through $360^\circ$ in the $(HK0)$ scattering plane. After standard time-of-flight reduction, the data were corrected for sample-environment scattering, residual temperature-independent background, and the crystallographic symmetry of the kagome plane. A representative sequence is shown in Fig.~\ref{fig:si_data_processing}. The measured intensities were normalized using a vanadium standard, placing the neutron scattering data on an absolute intensity scale.

The first correction accounted for scattering from the sample environment,
including the helium exchange gas and the CYTOP used in the sample
assembly. An empty-holder measurement collected under the same instrumental
conditions was subtracted from the sample data. Because the ratio of sample
mass to CYTOP was large for this assembly, a single sample-environment
background measurement was sufficient. The resulting intensity is shown in
Fig.~\ref{fig:si_data_processing}(b).

A residual background remained after this subtraction, most visibly as
arc-like features at low energy transfer. These features arise primarily
from elastic Bragg scattering extending into the inelastic channel because
of the finite instrumental energy resolution. We removed this contribution
using the detailed-balance subtraction introduced in
Refs.~\cite{Helton2007,Breidenbach2025}. The method uses the
fluctuation-dissipation relation
\begin{equation}
S(\mathbf{Q},\omega,T)
=
\left[1+n(\omega,T)\right]
\chi''(\mathbf{Q},\omega,T),
\end{equation}
where $n(\omega,T)$ is the Bose occupation factor, together with
\begin{equation}
\chi''(\mathbf{Q},-\omega,T)
=
-\chi''(\mathbf{Q},\omega,T).
\end{equation}
CNCS measures both positive-energy-transfer events, in which the neutron
transfers energy to the sample, and negative-energy-transfer events, in
which the neutron gains energy from the sample. We write the measured
intensity as
\begin{equation}
\widetilde{S}(\mathbf{Q},\omega,T)
=
S(\mathbf{Q},\omega,T)+b(\mathbf{Q},\omega),
\end{equation} 
where $b(\mathbf{Q},\omega)$ is assumed to be temperature independent. Combining the positive- and negative-energy-transfer data measured at two temperatures allows the physical scattering and the temperature-independent background to be separated using detailed balance. For both incident energies, the corrected $T=1.7$ and 40~K spectra were extracted using the paired measurements at 1.7 and 40~K. For the $E_i=3.32$~meV data, the corrected 150~K spectrum was instead extracted using the paired measurements at 1.7 and 150~K. Because the elastic Bragg intensities and their resolution tails vary only weakly over the temperature range used for the subtraction, this procedure strongly suppresses the arc-like features while retaining the temperature-dependent inelastic response, as shown in Fig.~\ref{fig:si_data_processing}(c).

Finally, each constant-energy slice was symmetrized over the $D_6$-equivalent momenta of the kagome plane. Intensities at symmetry-related points were averaged to improve counting statistics and produce the spectra shown in Fig.~\ref{fig:si_data_processing}(d). Unless otherwise specified, the experimental analyses and comparisons with the numerical calculations were performed using these fully corrected and $D_6$-symmetrized data.

\subsection{Temperature dependence and phonon contributions to the
neutron scattering}

\begin{figure*}[htbp]
\centering
\includegraphics[width=0.95\textwidth]{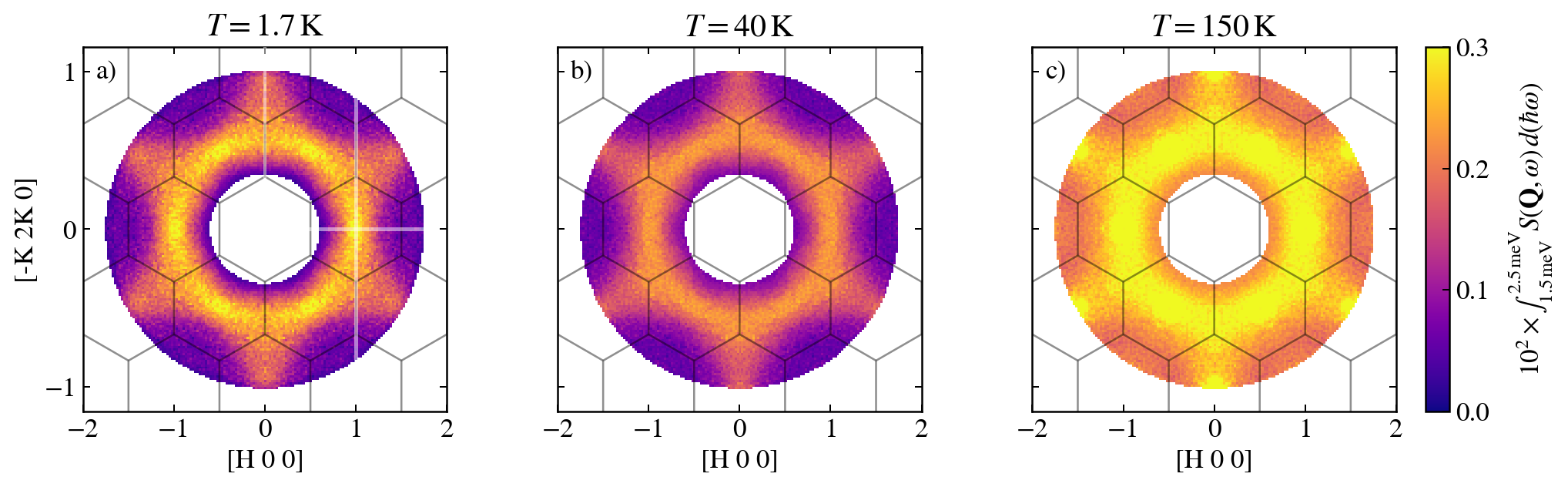}
\caption{\textbf{Temperature dependence of the scattering in the
$(HK0)$ plane.}
Neutron scattering integrated over $1.5 \leq \hbar\omega \leq 2.5$~meV at (a) $T=1.7$~K, (b) $T=40$~K, and (c) $T=150$~K. All panels use the same intensity scale. Gray lines indicate Brillouin-zone boundaries. The white lines in panel (a) mark the momentum trajectories shown in Fig.~\ref{fig:si_energy_momentum_cuts}. The broad momentum dependence observed at low temperature is primarily magnetic, while the sharper temperature-enhanced features near $(100)$- and $(110)$-type reciprocal-lattice positions are assigned to phonon scattering.}
\label{fig:si_hk0_temperature}
\end{figure*}

Figure~\ref{fig:si_hk0_temperature} shows constant-energy slices in the $(HK0)$ plane integrated over $1.5 \leq \hbar\omega \leq 2.5$~meV. This energy interval lies above the impurity-dominated low-energy response identified in the main text and is therefore predominantly sensitive to the intrinsic kagome response and phonon scattering. All three temperatures are plotted on the same intensity scale.

At $T=1.7$~K, the scattering forms the broad ring-like pattern characteristic of short-range kagome spin correlations. Its overall intensity decreases modestly between 1.7 and 40~K, while the principal momentum dependence remains visible. At 150~K, sharper intensity enhancements associated with the $(110)$ phonon become prominent. Away from these localized phonon features, a broad momentum-dependent response remains visible. Because 150~K is still below the magnitude of the measured Curie-Weiss temperature, $|\Theta_{\mathrm{CW}}|\simeq253$~K, the persistence of this diffuse response is consistent with substantial short-range antiferromagnetic correlations well above the low-temperature QSL regime.

\begin{figure*}[htbp]
\centering
\includegraphics[width=0.95\textwidth]{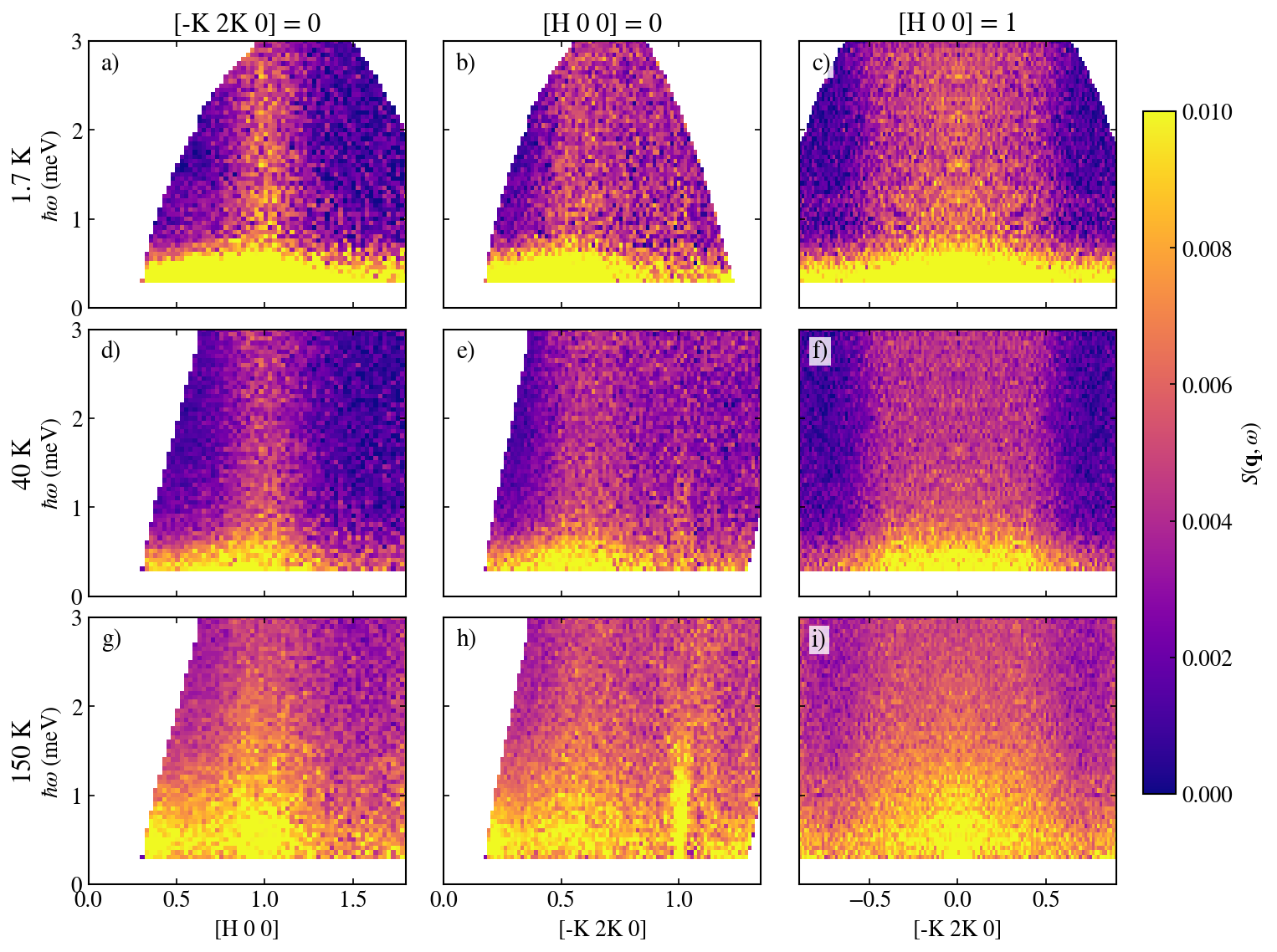}
\caption{\textbf{Energy-momentum slices at three temperatures.} Neutron scattering measured at $T=1.7$~K (a--c), $T=40$~K (d--f), and $T=150$~K (g--i). The left column shows scans along $[H\,0\,0]$ at $[-K\,2K\,0]=0$. The center column shows scans along $[-K\,2K\,0]$ at $[H\,0\,0]=0$, and the right column shows scans along $[-K\,2K\,0]$ at $[H\,0\,0]=1$. All panels use the same intensity scale. The broad, weakly dispersive response is associated primarily with magnetic scattering, whereas the sharper features that increase strongly with temperature, including the feature near the $(110)$ position, are assigned to phonons.}
\label{fig:si_energy_momentum_cuts}
\end{figure*}

The temperature evolution can be seen more clearly in the energy-momentum slices in Fig.~\ref{fig:si_energy_momentum_cuts}. The three columns follow the momentum trajectories indicated by the white lines in Fig.~\ref{fig:si_hk0_temperature}(a), and the rows show the same cuts at 1.7, 40, and 150~K.

\begin{figure*}[htbp]
\centering
\includegraphics[width=0.95\textwidth]
{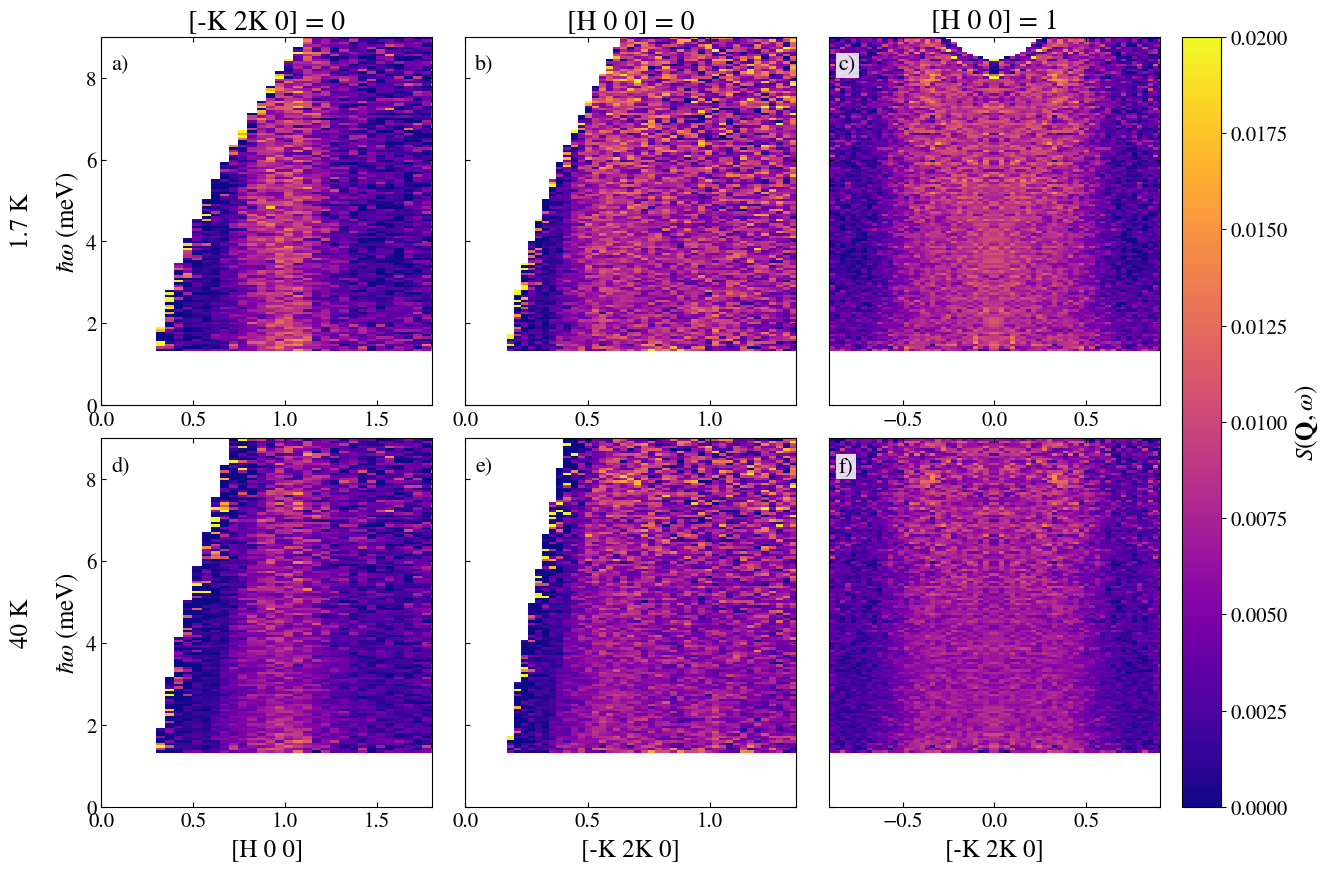} \caption{\textbf{Higher-energy magnetic scattering.} Energy-momentum slices measured with $E_i=12$~meV at $T=1.7$~K (a--c) and $T=40$~K (d--f). The left column shows scans along $[H\,0\,0]$ at $[-K\,2K\,0]=0$, the center column shows scans along $[-K\,2K\,0]$ at $[H\,0\,0]=0$, and the right column shows scans along $[-K\,2K\,0]$ at $[H\,0\,0]=1$. All panels use the same intensity scale. The region below $\hbar\omega=1.3$~meV is masked to emphasize the higher-energy response. The broad, weakly dispersive scattering remains visible over much of the accessible energy range at both temperatures, showing that the magnetic spectral weight extends well above the gap-like onset near 1~meV.}\label{fig:high_energy_magnetic} 
\end{figure*}

The contrasting temperature dependences help distinguish the two contributions. For positive energy transfer, one-phonon intensity is enhanced approximately by the Bose population factor $n(\omega,T)+1$. The sharp feature near $(110)$ increases strongly on warming and follows this expected behavior. The broad magnetic response behaves differently. In particular, its measured intensity is slightly reduced between 1.7 and 40~K even though the Bose factor increases substantially over this temperature range. The corresponding intrinsic magnetic susceptibility $\chi''(\mathbf{Q},\omega)$ must therefore decrease with temperature. The broad response nevertheless remains visible at 150~K, consistent with short-range magnetic correlations persisting to temperatures of order $J_1/k_{\mathrm B}$. The higher-incident-energy measurements show that the broad magnetic response extends well beyond the energy range accessible with $E_i=3.32$ meV. Figure~\ref{fig:high_energy_magnetic} shows representative energy-momentum slices measured with $E_i=12$ meV at 1.7 and 40 K. The broad, weakly dispersive scattering remains visible to high energies at both temperatures, demonstrating that the magnetic spectral weight extends over a broad energy range.

\begin{figure*}[htbp]
\centering
\includegraphics[width=0.6\textwidth]{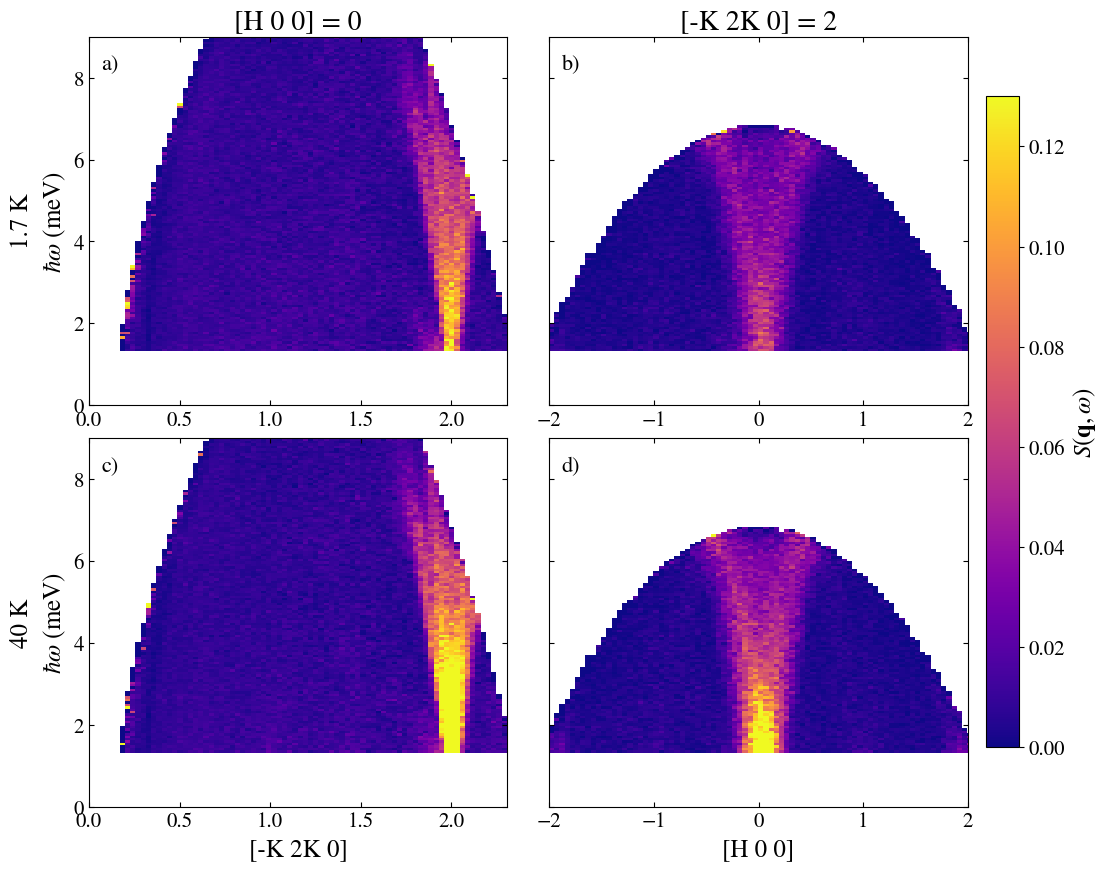}
\caption{\textbf{Directional dependence of the phonon near $(220)$.}
Energy-momentum slices through the phonon near the $(220)$
reciprocal-lattice position at $T=1.7$~K (a,b) and $T=40$~K
(c,d). Panels (a,c) show the dispersion along $[-K\,2K\,0]$ at
$[H\,0\,0]=0$, while panels (b,d) show the dispersion along
$[H\,0\,0]$ at $[-K\,2K\,0]=2$. The data indicate an apparent
two-ridge structure along $[-K\,2K\,0]$, whereas the corresponding
feature along $[H\,0\,0]$ is consistent with a single dominant
dispersion. All panels use the same intensity scale. The cuts are shown without $D_6$ symmetrization in order to preserve
the directional information.}
\label{fig:si_220_phonon}
\end{figure*}

Energy-momentum slices measured with $E_i=12$~meV through the phonon near the $(220)$ reciprocal-lattice position in the higher-incident-energy measurements, as shown in Fig.~\ref{fig:si_220_phonon}. These data also reveal a possible directional anisotropy of the phonon spectrum.

Along $[H\,0\,0]$, the $(220)$ feature appears as a single dominant
dispersive branch. Along $[-K\,2K\,0]$, two intensity ridges can be
resolved over part of the measured energy range. Notably, this
directional distinction coincides with the bond-direction anisotropy
of the chevron QSL shown in Fig.~1(d) of the main text, where
alternating strong and weak spin correlations break the rotational
symmetry of the kagome lattice. If this bond anisotropy couples to
lattice displacements, it could lift a phonon degeneracy or redistribute
phonon spectral weight differently along inequivalent momentum
directions. The observed phonon anisotropy is therefore qualitatively
compatible with chevron-like correlations.

\subsection{Pair-correlation model}

The pair-correlation model (PCM) expresses the momentum dependence of the
energy-integrated magnetic scattering in terms of real-space spin
correlations \cite{Han2016,Breidenbach2025}. Writing $E=\hbar\omega$, the
form-factor-corrected intensity can be approximated as
\begin{equation}
\widetilde{S}(\mathbf{Q})
\equiv
\frac{1}{|F(\mathbf{Q})|^2}
\int_0^{E^\ast} dE\, I(\mathbf{Q},E)
\simeq
\alpha
\left[
1+\sum_{k=1}^{p}
m_k g_k(\mathbf{Q}) C_k
\right],
\label{eq:si_pcm}
\end{equation}
where
\begin{equation}
C_k =
\frac{
\langle \mathbf{S}_i\cdot\mathbf{S}_j\rangle_k
}{
\langle S^2\rangle
},
\qquad
g_k(\mathbf{Q}) =
\frac{1}{|\mathcal{R}_k|}
\sum_{\mathbf{r}\in\mathcal{R}_k}
\cos(\mathbf{Q}\cdot\mathbf{r}).
\end{equation}
Here, $C_k$ is the average correlation for the $k$th pair shell,
$m_k$ is its coordination number, and $\mathcal{R}_k$ contains the
symmetry-equivalent bond vectors in that shell. The normalization $\alpha$
accounts for the overall intensity scale and the finite energy-integration
window.

For the analysis in the main text, we use $E^\ast=8$~meV. Extending the
integration to higher energies substantially reduces the available momentum
coverage because of kinematic constraints. Equation~\ref{eq:si_pcm} is
linear in $\alpha$ and $\alpha m_k C_k$, allowing the correlations to be
obtained by linear least-squares regression over the measured momentum
points. Counting-statistics uncertainties are propagated through the
regression assuming independent errors between momentum bins.

The model contains 15 kagome pair-correlation terms extending to
$d=6$ nearest-neighbor spacings. The five shortest impurity-related pair
terms are also included so that their momentum dependence is not absorbed
into the fitted kagome correlations. The finite momentum coverage limits
the number of pair terms that can be distinguished reliably, since different
$g_k(\mathbf{Q})$ can become strongly correlated over the measured region.

\begin{figure*}[htbp]
\centering
\includegraphics[width=0.55\textwidth]{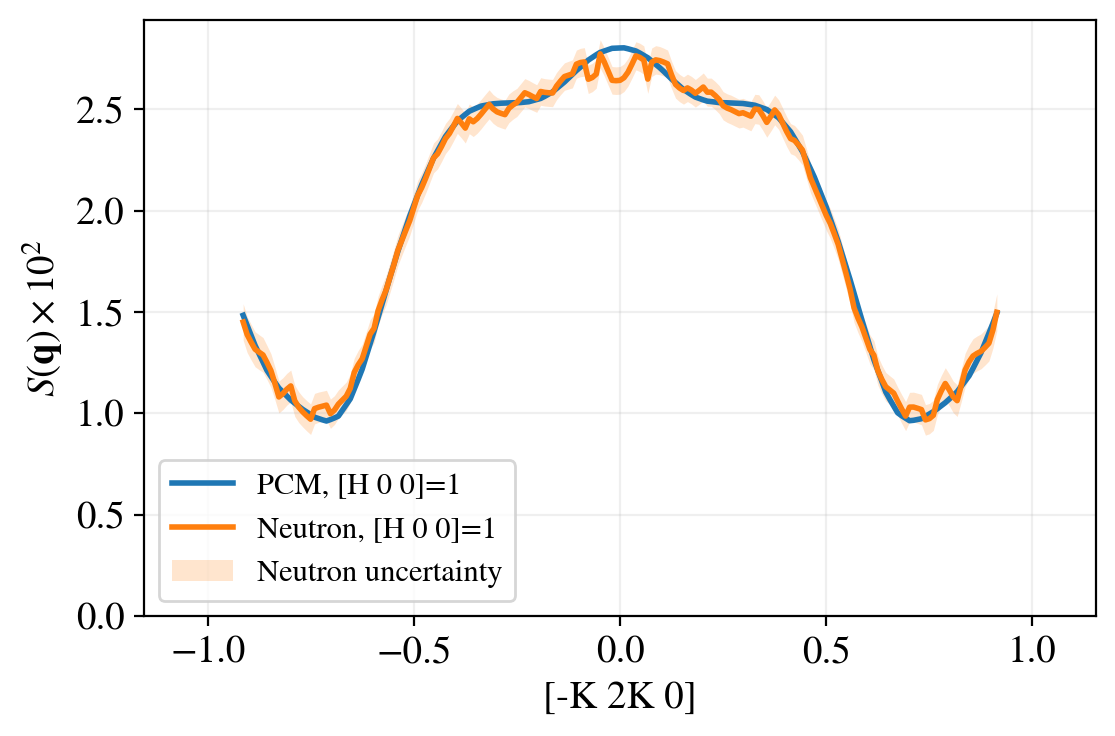}
\caption{\textbf{Line cut through the pair-correlation reconstruction at
$T=1.7$~K.}
Measured neutron intensity integrated from 0 to 8~meV and the corresponding
PCM reconstruction along $[H\,0\,0]=1$, plotted as a function of
$[-K\,2K\,0]$. The shaded region shows the neutron counting uncertainty.
The PCM reproduces the broad maxima and minima of the measured structure
factor within the experimental uncertainty.}
\label{fig:si_pcm_linecut}
\end{figure*}

Figure~\ref{fig:si_pcm_linecut} shows a representative line cut through the
measured and reconstructed structure factors. The model follows the detailed
momentum dependence of the data, including the minima near
$[-K\,2K\,0]\simeq\pm0.7$ and the broad maximum around the zone center.

\begin{figure*}[htbp]
\centering
\includegraphics[width=0.95\textwidth]{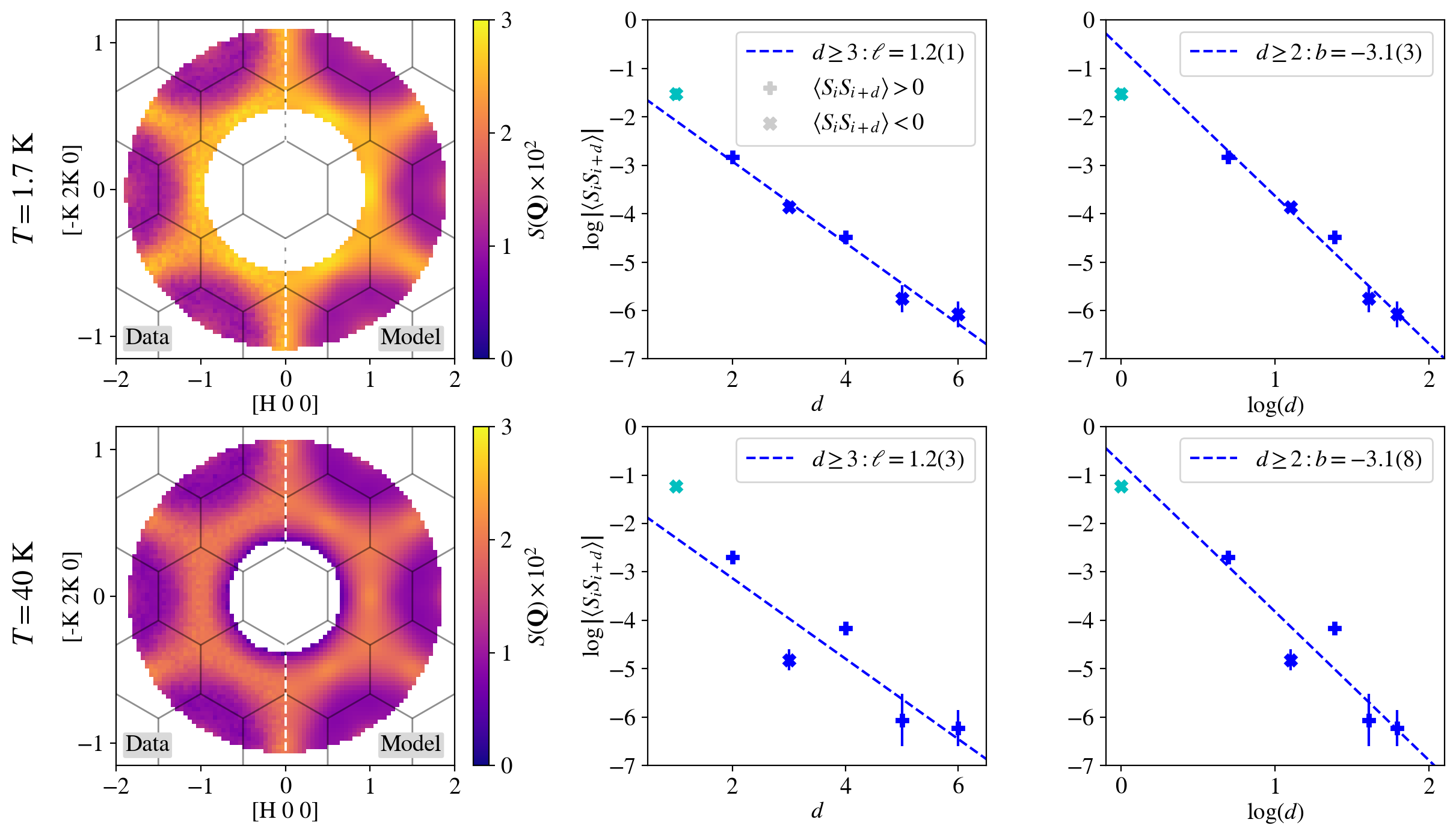}
\caption{\textbf{Pair-correlation analysis at $T=1.7$~K and $T=40$~K.}
The top and bottom rows show results at $T=1.7$~K and $T=40$~K,
respectively. Left column: measured intensity on the left half and PCM
reconstruction on the right half, integrated from 0 to 8~meV. Middle column:
$\log|\langle\mathbf{S}_i\cdot\mathbf{S}_{i+d}\rangle|$ plotted against
distance $d$ in units of the nearest-neighbor spacing. Dashed lines show
exponential fits, $|C(d)|\propto\exp(-d/\ell)$, for $d\geq2$. Right column:
the same correlations plotted against $\log d$, with power-law fits
$|C(d)|\propto d^b$ for $d\geq2$. Marker shape distinguishes positive and
negative correlations. The exponential fits give
$\ell=1.2(1)$ and $1.2(3)$ at 1.7 and 40~K, respectively, while the
power-law fits give $b=-3.1(3)$ and $-3.1(8)$.}
\label{fig:si_pcm_temperature}
\end{figure*}

Repeating the analysis at 40~K changes several of the individual
further-neighbor correlations, but the overall decay scale remains similar.
The exponential and power-law parameters at the two temperatures agree
within their uncertainties. Because the experimentally accessible range
extends only to $d=6$, both functional forms provide reasonable empirical
descriptions of the decay though the data is more consistent with the exponential decay. These fits therefore quantify its rapid spatial
suppression but do not by themselves determine its asymptotic functional
form.


\subsection{Modeling of gap phenomenon}

\begin{figure}[htbp]
\includegraphics[width=0.65\textwidth]{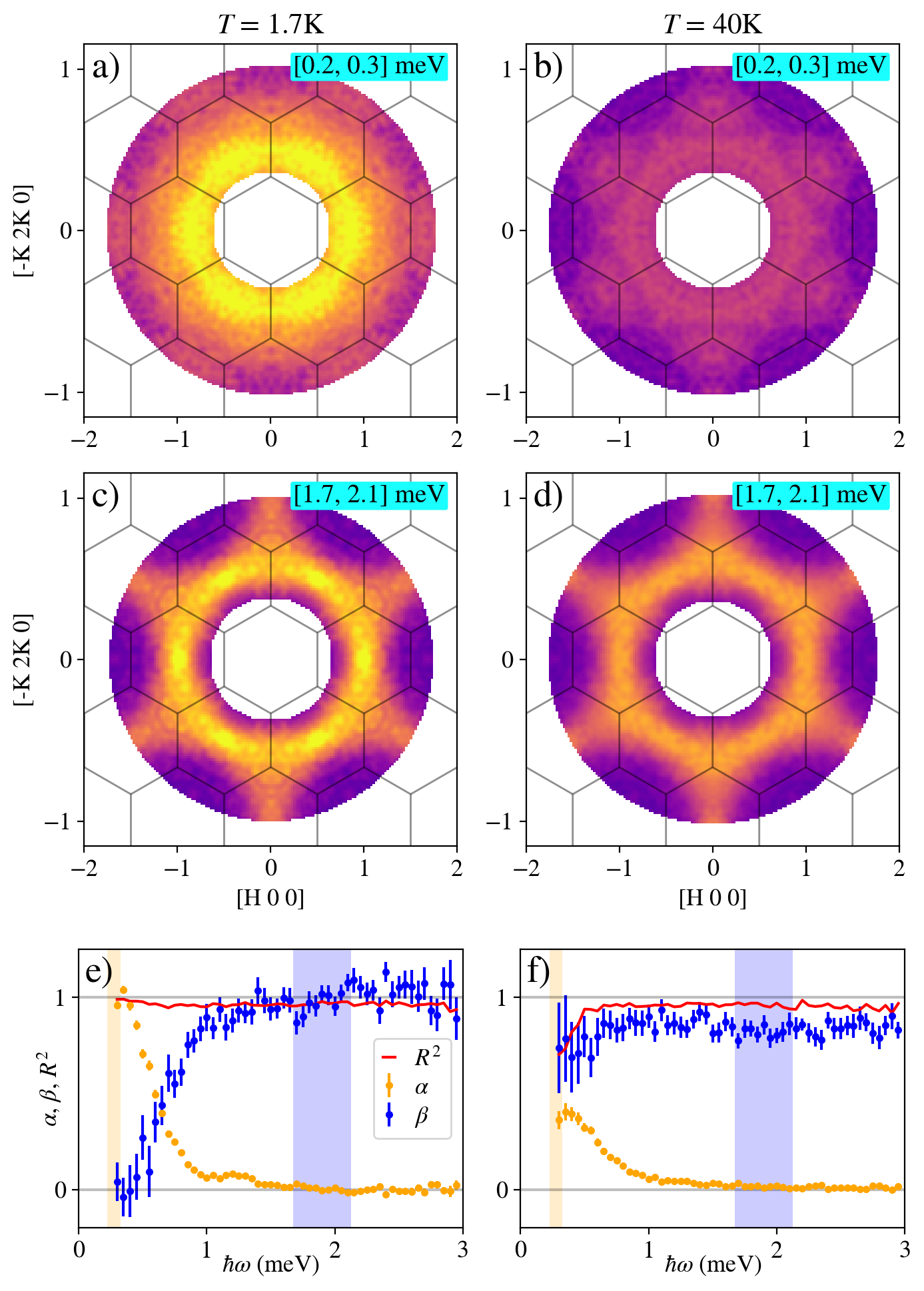}
\caption{\label{fig:si_two_slice}
\textbf{Temperature dependence of the two-component decomposition.}
\textbf{(a,b)} Scattering contours integrated over $0.2\leq\hbar\omega\leq0.3$~meV at $T=1.7$ and $40$~K, respectively.
\textbf{(c,d)} Scattering contours integrated over $1.7\leq\hbar\omega\leq2.1$~meV at $T=1.7$ and $40$~K, respectively. Panels (a,b) share one intensity scale, and panels (c,d) share one intensity scale.
\textbf{(e)} Fitted coefficients $\alpha$ and $\beta$ and coefficient of determination $R^2$ obtained by modeling the $T=1.7$~K data using the low- and high-energy maps in panels (a) and (c) as basis functions.
\textbf{(f)} Corresponding fit of the $T=40$~K data using the same base-temperature basis functions. Shaded regions mark the energy windows used to define the basis slices.}
\end{figure}

Figure~\ref{fig:si_two_slice} extends the two-component decomposition in the main text to $T=40$~K. The low- and high-energy maps measured at base temperature are used as fixed basis functions, and each energy slice at $T=1.7$ and $40$~K is modeled as
\[
I(\mathbf{Q},\omega')=
\alpha I(\mathbf{Q},\omega_\mathrm{Low})
+\beta I(\mathbf{Q},\omega_\mathrm{High}).
\]
At base temperature, $R^2$ remains high over most of the measured energy range, while the impurity-associated coefficient $\alpha$ decreases rapidly and the high-energy coefficient $\beta$ rises through the gap-like onset. Applying the same basis functions to the $40$~K data captures the broad evolution of the momentum dependence, although the fit quality is reduced, particularly below approximately $0.5$~meV. The impurity-associated component is suppressed, while the high-energy component acquires weight at lower energies. Thus, warming changes the relative spectral weights and cannot be represented as a simple overall rescaling of the base-temperature response.

To characterize the momentum dependence of the gap-like onset, we extracted energy cuts from a grid of momentum-space regions in the $(HK0)$ plane. Each region includes all symmetry-equivalent momenta related by the $D_6$ point group, improving the counting statistics while imposing the approximate sixfold symmetry observed in the data. The cuts were converted from $S(\mathbf{Q},\omega)$ to $\chi''(\mathbf{Q},\omega)$ using the fluctuation-dissipation relation and fitted between 0.3 and 2.5~meV in 0.05~meV energy intervals.

Following the phenomenological decomposition used for herbertsmithite and Zn-barlowite in Refs.~\cite{Han2016,Breidenbach2025}, each cut was modeled as the sum of a damped harmonic oscillator (DHO) and a sigmoid:
\begin{equation}
\chi''(\mathbf{Q},E)
=
A_{\mathrm{DHO}}
\frac{2E\lambda}
{\left(\omega_0^2-E^2\right)^2+E^2\lambda^2}
+
\frac{A_{\mathrm{K}}}
{1+\exp[-s(E-E_{\mathrm{gap}})]},
\label{eq:si_dho_sigmoid}
\end{equation}
where $E=\hbar\omega$ is the energy transfer. The DHO describes the impurity-dominated low-energy response. Its amplitude, damping, and resonance energy are denoted by $A_{\mathrm{DHO}}$, $\lambda$, and $\omega_0$, respectively. The sigmoid describes the onset of the higher-energy kagome response, with amplitude $A_{\mathrm{K}}$, steepness $s$, and midpoint $E_{\mathrm{gap}}$. Thus, $E_{\mathrm{gap}}$ is the energy at which the fitted sigmoid reaches half of its asymptotic amplitude. We refer to this quantity as the gap-like onset energy. It should not by itself be interpreted as a direct measurement of the many-body spin gap of the clean kagome Hamiltonian.

We first performed an unconstrained fit in which all six parameters in Eq.~\ref{eq:si_dho_sigmoid} were allowed to vary independently at each momentum. These fits provide a direct test of whether the characteristic energy scales of the impurity response vary across momentum. However, the limited energy range and partial overlap of the two components produce correlations among $\lambda$, $\omega_0$, $s$, and $E_{\mathrm{gap}}$. Consequently, the unconstrained fits can become poorly conditioned in regions with lower intensity.

For the constrained analysis used in the main text, we fixed the DHO damping and resonance energy to
\begin{equation}
\lambda = 0.617~\mathrm{meV},
\qquad
\omega_0 = 0.469~\mathrm{meV}.
\end{equation}
These values were obtained from an unconstrained fit to the momentum-integrated herbertsmithite spectrum, which gave $\lambda=0.617(19)$~meV and $\omega_0=0.469(3)$~meV. The DHO amplitude and all three sigmoid parameters were then fitted independently for each momentum region. This procedure assumes that the characteristic line shape of the impurity mode is approximately momentum independent while allowing its spectral weight, as well as the intrinsic onset energy, to vary with momentum. Reducing the number of free parameters substantially stabilizes the extraction of $E_{\mathrm{gap}}$.

The best-fit parameters were obtained by nonlinear least-squares minimization. The reported parameter uncertainties are one-standard-deviation errors calculated from the square roots of the diagonal elements of the fit covariance matrix. The fits used equal weighting of the energy points. The resulting errors therefore describe the local statistical uncertainty of the nonlinear fit as estimated from its residuals and parameter curvature. They do not include possible systematic errors associated with the choice of fitting function, momentum-mask size, fitting window, or the assumption of a momentum-independent DHO line shape. In the constrained fits, the uncertainties in the fixed values of $\lambda$ and $\omega_0$ are also not propagated into the quoted uncertainty of $E_{\mathrm{gap}}$.

\begin{figure*}[htbp]
\includegraphics[width=0.95\textwidth]{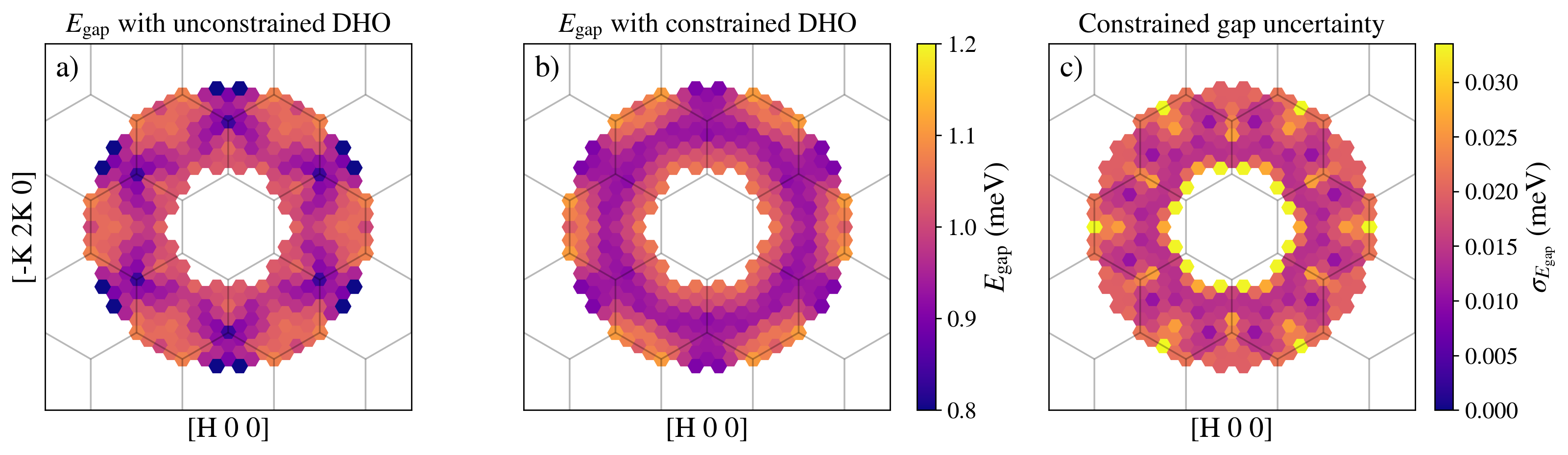}
\caption{
\textbf{Momentum dependence and uncertainty of the gap-like onset energy in herbertsmithite.}
\textbf{Left:} Gap-like onset energy $E_{\mathrm{gap}}$ obtained from the fully unconstrained DHO plus sigmoid fits, in which the DHO amplitude, damping $\lambda$, resonance energy $\omega_0$, sigmoid amplitude, steepness, and onset energy were all allowed to vary independently. Points with $\sigma_{E_{\mathrm{gap}}}/E_{\mathrm{gap}}>0.30$ are omitted.
\textbf{Center:} $E_{\mathrm{gap}}$ obtained from the constrained fits used in the main text, with the DHO damping and resonance fixed to $\lambda=0.617$~meV and $\omega_0=0.469$~meV, respectively.
\textbf{Right:} One-standard-deviation uncertainty $\sigma_{E_{\mathrm{gap}}}$ of the constrained-fit onset energy, calculated from the diagonal of the nonlinear-fit covariance matrix.
Each fitted value is repeated at the corresponding $D_6$-equivalent momenta. Gray lines indicate the Brillouin-zone boundaries. The left and center panels use the same color scale.
}
\label{fig:si_gap_fit_maps}
\end{figure*}

Figure~\ref{fig:si_gap_fit_maps} compares the onset energies obtained from the unconstrained and constrained analyses. The shared color scale permits a direct comparison of the momentum dependence. The unconstrained results reproduce the broad variation found in the constrained analysis, although some momenta have larger uncertainties because of correlations among the six free parameters. The constrained fit produces a more stable estimate of the gap-like onset over the accessible momentum range. The corresponding uncertainty map shows that the momentum-dependent modulation is generally larger than the local fitting uncertainty. Averaged over the momenta included in the main-text figure, the constrained analysis gives $E_{\mathrm{gap}}=0.995$~meV with a mean fit uncertainty of 0.024~meV.

\begin{figure*}[htbp]
\includegraphics[width=0.65\textwidth]{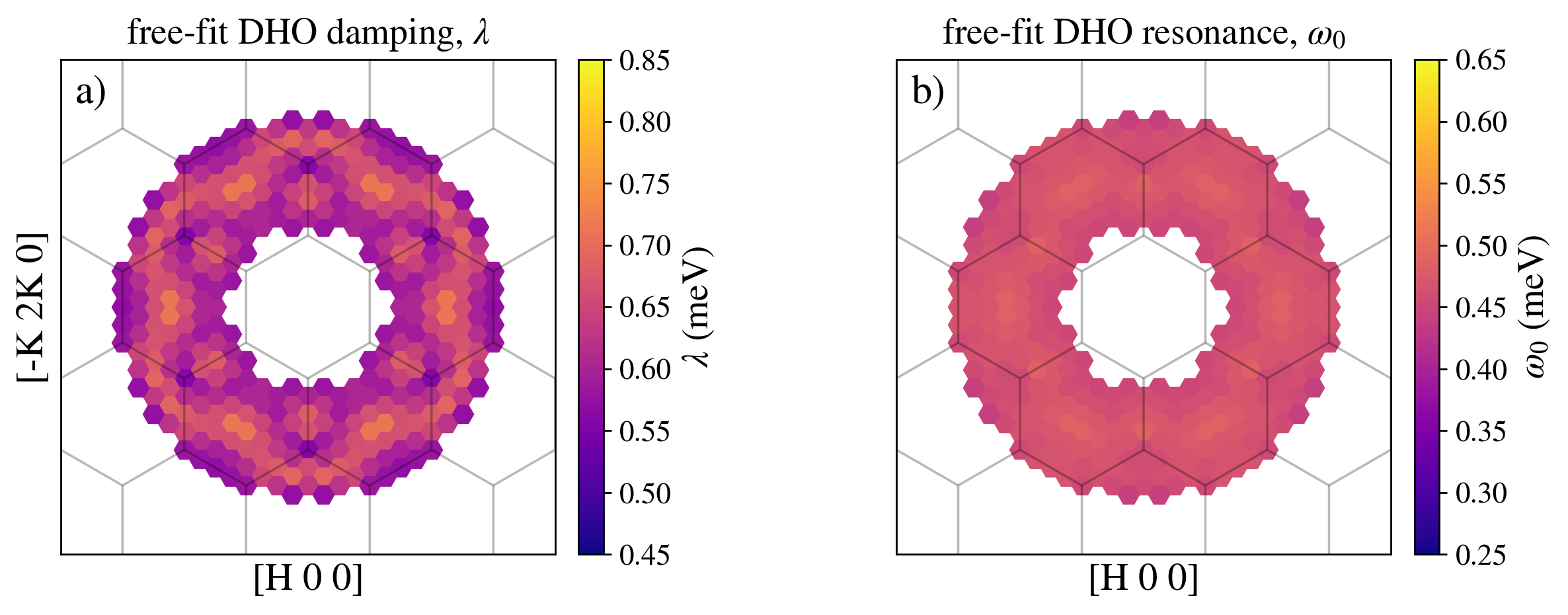}
\caption{
\textbf{Momentum dependence of the unconstrained DHO parameters in herbertsmithite.}
\textbf{Left:} DHO damping parameter $\lambda$ obtained by allowing all six parameters of the DHO plus sigmoid model to vary independently.
\textbf{Right:} Corresponding DHO resonance energy $\omega_0$.
Only points for which the one-standard-deviation uncertainty is no greater than 30\% of the fitted parameter value are shown. The uncertainties are obtained from the diagonal elements of the nonlinear-fit covariance matrix. Each value is repeated at its $D_6$-equivalent momenta, and gray lines show the Brillouin-zone boundaries. Separate color scales are used because $\lambda$ and $\omega_0$ have different numerical ranges.
}
\label{fig:si_dho_parameters}
\end{figure*}

Figure~\ref{fig:si_dho_parameters} shows the independently fitted DHO damping and resonance energy from the unconstrained analysis. These maps provide a check on the use of common DHO parameters in the constrained fits. They also identify momentum regions in which the impurity line shape is weakly constrained by the available energy cut. For the unconstrained gap and DHO-parameter maps, points were retained only when the one-standard-deviation uncertainty of the plotted parameter was no greater than 30\% of its fitted value.

\clearpage
\section{Further details of DMRG \& DDMRG simulations}

\subsection{The XC cylinder and spin exchanges}

\begin{figure*}[h]
\includegraphics[width=0.6\textwidth]{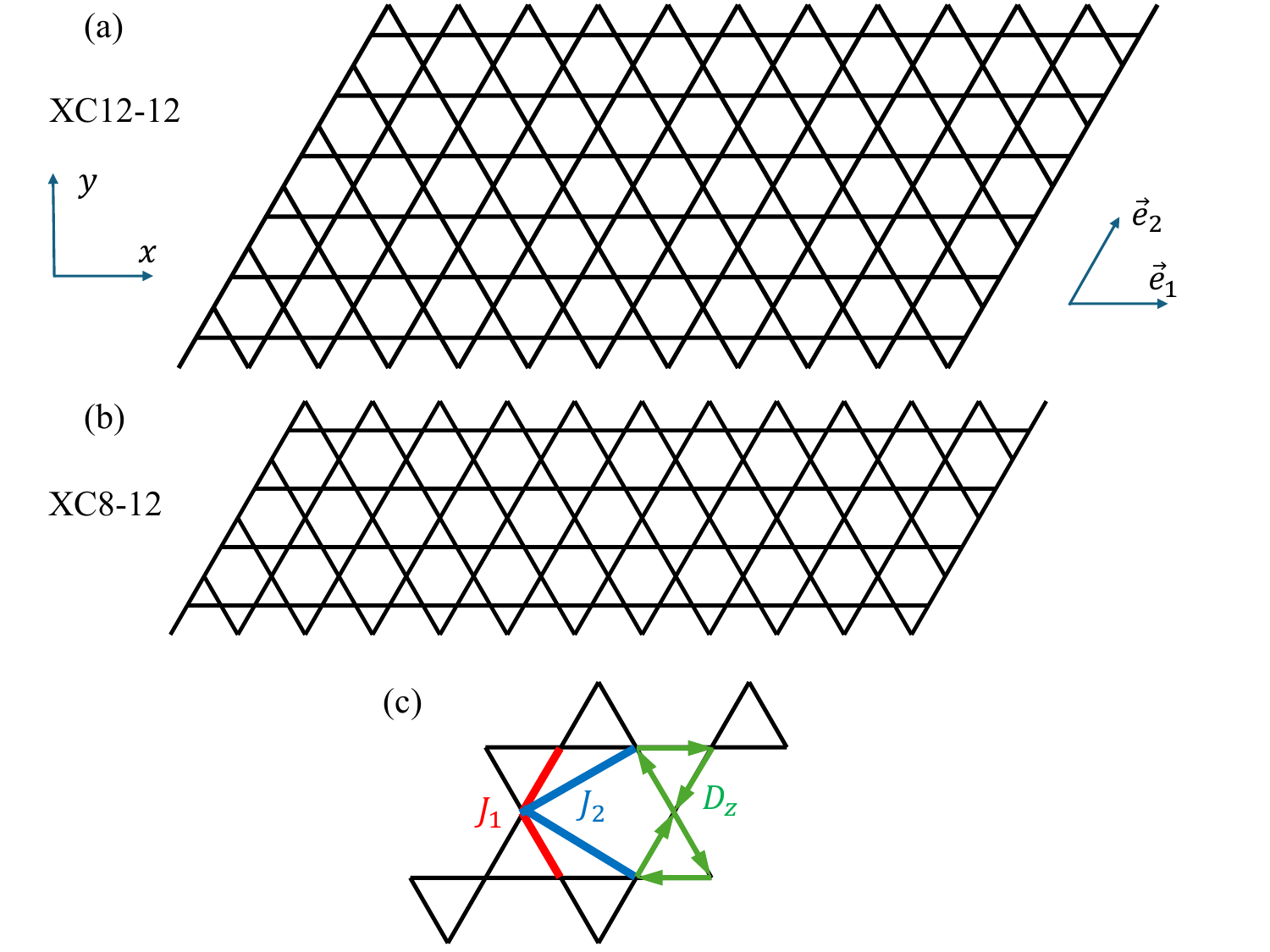}
\caption{\textbf{XC cylinders and spin exchanges} (a) The XC12-12 cluster with 12 unit cells along the $\vec{e}_1$ direction and 6 unit cells (12 sites) along the $\vec{e}_2$ direction. $\vec{e}_1$ and $\vec{e}_2$ denote the two unit vectors of the cylinder with open and periodic boundary conditions, respectively. (b) The XC8-12 cylinder. (c) The nearest-neighbor exchange $J_1$, second nearest-neighbor exchange $J_2$, and out-of-plane DM interactions $D_z$. The arrows on the bond define the sequence of spins under antisymmetric exchange $D_z$.}
\label{sifig:lattice}
\end{figure*}

In this section, we first introduce the cluster we used in the density-matrix renormalization group (DMRG) and dynamical-DMRG simulations.  Following the convention in Ref.~\cite{Yan2011,Kolley2015}, we refer to a cylinder with $L_x$ and $L_y$ unit cells ($2L_x$ and $2L_y$ sites) in the $\vec{e}_1$ and $\vec{e}_2$ directions as XC$2L_y$-$L_x$ cylinder. $\vec{e}_1$ and $\vec{e}_2$ denote the two unit vectors of the cylinder with open and periodic boundary conditions, respectively. Note that $\vec{e}_2$ has a $30^\circ$ angle to the $y$ direction in the Cartesian coordinate. We considered two cylinders in this paper, the XC12-12 cylinder shown in Fig.~\ref{sifig:lattice}(a), and the XC8-12 cylinder shown in Fig.~\ref{sifig:lattice}(b).

In Fig.~\ref{sifig:lattice}(c), we illustrate the various spin exchanges in the $S=1/2$ kagome Heisenberg model we study. The Hamiltonian is:
\begin{equation}
    \label{Eq:Ham}
H = J_1\sum_{\langle ij\rangle} \mathbf{S}_i\cdot \mathbf{S}_j + J_2\sum_{\langle \langle ij\rangle \rangle} \mathbf{S}_i\cdot \mathbf{S}_j
+ \sum_{\langle ij\rangle}\mathbf{D}_{ij}\cdot (\mathbf{S}_i\times \mathbf{S}_j ).
\end{equation},
where $J_1$ and $J_2$ are the Heisenberg exchange over first- and second-nearest-neighbor pairs of sites, respectively.
The third term is the Dzyaloshinskii-Moriya (DM) exchange originating from spin-orbit coupling, which is present as a result of broken inversion symmetry ~\cite{Dzyaloshinsky1958,Moriya1960,Lee2013}.
The DM interaction is anti-symmetric under exchange of $i$ and $j$~\cite{Elhajal2002,Cepas2008,Zorko2008}, so a convention of the spin exchanges needs to be specified. 
We follow the convention in Ref.~\cite{Lee2013} to arrange the bonds $i\rightarrow j$ in a clockwise way and choose $D_z>0$, as shown in Fig.~\ref{sifig:lattice}(c).
Since electron spin resonance measurements suggest the existence of a leading out-of-plane component of $\mathbf{D}_{ij}$~\cite{Zorko2008}, we hence simplify $\mathbf{D}_{ij}=D\hat{z}$ in the present study. 
This reduces the computational costs and facilitates reliable DMRG simulations on large systems by leveraging the $U(1)$ spin rotational symmetry.

\subsection{Equal-time spin structure factor $S(\mathbf{Q})$ without symmetrization or form factor}

Here we show the raw version of $S(\mathbf{Q})$ obtained directly from DMRG, whose symmetrized version is compared to that from the neutron after applying the form factor in the main text.
The $S(\mathbf{Q})$ of the isotropic QSL exhibits small peaks at some but not all $\Gamma'$ points, since the cylinder we use breaks $C_6$ rotational symmetry. The chevron QSL shows peaks around but not exactly at the $K'$ points. However, since the peaks are not related by $C_6$ symmetry, after symmetrization they average out and form a high-intensity region centered at the $K'$ points.

\begin{figure*}[h]
\vskip -0.3cm
\includegraphics[width=0.8\textwidth]{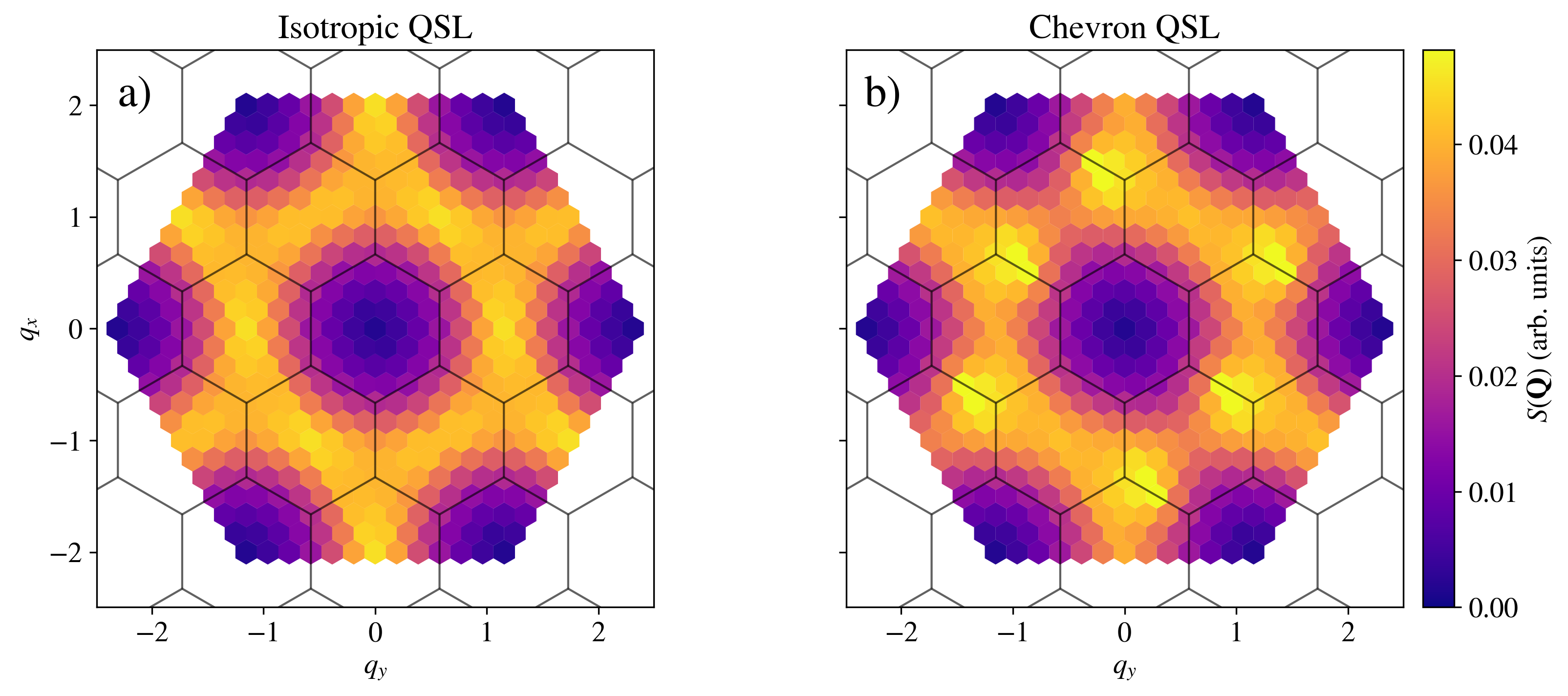}
\vskip -0.3cm
\caption{\textbf{Unsymmetrized $S(\mathbf{Q})$ from DMRG.} Unsymmetrized $S(\mathbf{Q})$ from DMRG simulation on XC12-12 cylinder for (a) the isotropic QSL and (b) the chevron QSL.}
\end{figure*}

\subsection{Simulating the dynamical spin structure factor $S(\mathbf{Q},\omega)$ with the dynamical-DMRG algorithm}

The goal is to calculate the dynamical spin structure factor (DSSF), defined as:
\begin{equation}
\begin{split}
S(\mathbf{Q},\omega)=&\frac{-1}{\pi N}\sum_{\substack{i,j=1\\\alpha=x,y,z}}^{N}e^{i\mathbf{Q}\cdot(\mathbf{r}_i-\mathbf{r}_j)}\times\\
&\text{Im}\langle 0|S^\alpha_{\mathbf{r}_i}\frac{1}{E_0+\omega-H+i\eta}S^\alpha_{\mathbf{r}_j}|0\rangle,
\label{eq:dssf}
\end{split}
\end{equation}
where $|0\rangle$ is the ground state obtained by DMRG that has energy $E_0$, and $\eta$ is the broadening factor. 
We define the correction vector $|c^\alpha_{\mathbf{r}_i}(\omega)\rangle$ that satisfies:
\begin{equation}
(E_0+\omega-H+i\eta)|c^\alpha_{\mathbf{r}_j}(\omega)\rangle=S^\alpha_{\mathbf{r}_j}|0\rangle, 
\end{equation}
the $S(\mathbf{Q},\omega)$ can then be computed with:
\begin{equation}
S(\mathbf{Q},\omega)=\frac{-1}{\pi N}\sum_{\substack{i,j=1\\\alpha=x,y,z}}^{N}e^{i\mathbf{Q}\cdot(\mathbf{r}_i-\mathbf{r}_j)}\times\text{Im}\langle 0|S^\alpha_{\mathbf{r}_i}|c^\alpha_{\mathbf{r}_j}(\omega)\rangle.
\end{equation}
The correction vector is separated into real and imaginary parts: $|X^\alpha_{\mathbf{r}_j}(\omega)\rangle=\text{Re}|c^\alpha_{\mathbf{r}_j}(\omega)\rangle$ and $|Y^\alpha_{\mathbf{r}_j}(\omega)\rangle=\text{Im}|c^\alpha_{\mathbf{r}_j}(\omega)\rangle$, so that it can be solved with real arithmetic. Its imaginary part satisfies:
\begin{equation}
    [(E_0+\omega-H)^2+\eta^2]|Y^\alpha_{\mathbf{r}_j}(\omega)\rangle=-\eta S^\alpha_{\mathbf{r}_j}|0\rangle, 
    \label{eq:ddmrg}
\end{equation}
based on which the real part can also be computed:
\begin{equation}
    |X^\alpha_{\mathbf{r}_j}(\omega)\rangle=\frac{H-E_0-\omega}{\eta}|Y^\alpha_{\mathbf{r}_j}(\omega)\rangle.
\end{equation}
The dynamical-DMRG algorithm~\cite{ddmrg-jeckelmann} solves Eq.~\ref{eq:ddmrg} using the same sweep scheme as the DMRG algorithm~\cite{White1992,White1993}, replacing the original eigenvalue problem with the linear problem in Eq.~\ref{eq:ddmrg}. At each step of the sweep, an effective superblock Hamiltonian $H^{\rm eff}$ is formed by projecting the original Hamiltonian on the reduced basis. $H^{\rm eff}$ approximates the $H$ in Eq.~\ref{eq:ddmrg}, and this linear equation is solved using the conjugate gradient method. To minimize the effect of this approximation, $|X^\alpha_{\mathbf{r}_j}(\omega)\rangle$ needs to be state-averaged with $|Y^\alpha_{\mathbf{r}_j}(\omega)\rangle$ in DDMRG. 

Following Ref.~\cite{ddmrg-trij1j2}, we implemented several modifications to the original DDMRG algorithm, such that the correct qualitative features can be captured at a modest bond dimension with stable convergence.
The modifications are:  (1) use a separate MPS to store the ground state $|0\rangle$ and $S|0\rangle$ (following Ref.~\cite{ddmrg++}), mix the latter with a weight $\sqrt \eta$ in the state-averaged MPS, to mitigate the projection error when constructing the reduced linear problem (Eq.~\ref{eq:ddmrg}); (2) impose a weight $\eta$ on $|X(\omega)\rangle$ to de-emphasize it during state-averaging, which can preserve the accuracy of $|Y(\omega)\rangle$ that can help stabilize the algorithm when the bond dimension is relatively small. In this way, three states ($\sqrt \eta S^\alpha_{\vec{r}_b}|0\rangle$, $\eta|X(\omega)\rangle$, and $|Y(\omega)\rangle$) are state-averaged in one MPS, with the latter two being optimized during DDMRG.

In this paper, we select an $\eta=0.05J_1$ as the broadening factor. Because the $SU(2)$ symmetry is preserved in the QSL state we study, we replace the summation over spin component $\alpha$ in Eq.~\ref{eq:dssf} by the $z$ component multiplied by three.
To reduce boundary effects, the DDMRG simulations are carried out on the XC8-12 cylinder in Fig.~\ref{sifig:lattice}(b), and the Fourier transform in Eq.~\ref{eq:dssf} uses spin correlations from the central $4\times4$ unit cells (of size XC8-4).

\subsection{Convergence of $S(\mathbf{Q},\omega)$ from DDMRG}

\begin{figure*}[htbp]
\centering
\includegraphics[width=0.6\textwidth]{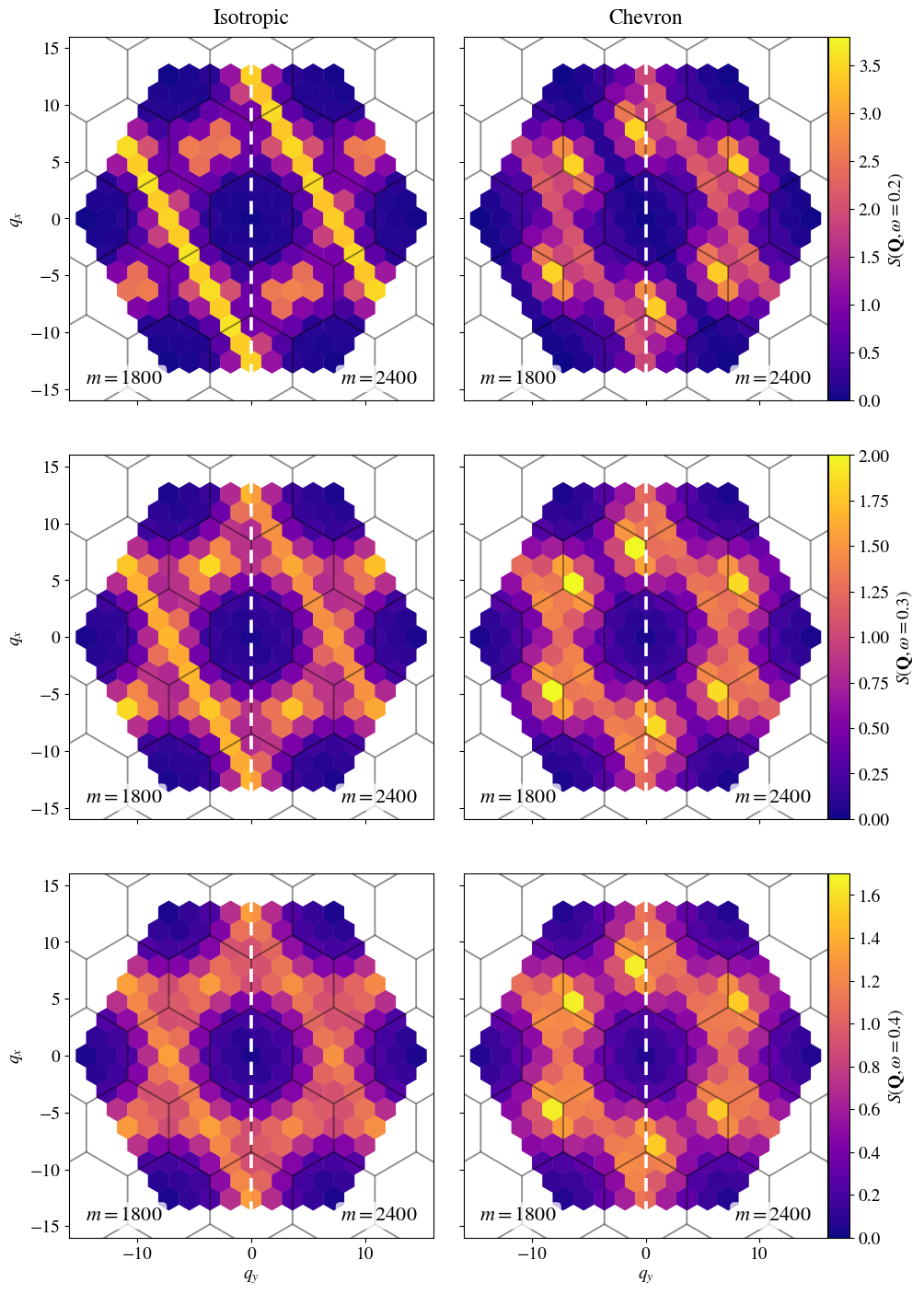}
\vskip-0.5cm
\caption{\textbf{Bond-dimension convergence of the DDMRG dynamical structure factor $S(\mathbf{Q},\omega)$.}
Unsymmetrized $S(\mathbf{Q},\omega)$ for the isotropic QSL
(left column) and chevron QSL (right column) at
$\omega=0.2J_1$, $0.3J_1$, and $0.4J_1$ from top to bottom.
Within each panel, the left half was calculated with bond dimension
$m=1800$ and the right half with $m=2400$. Each row uses a common
intensity scale.}
\label{sifig:convergence}
\end{figure*}

Figure~\ref{sifig:convergence} compares the unsymmetrized DDMRG spectra obtained with bond dimensions $m=1800$ and $m=2400$. Because the calculations are performed on cylinders, the raw spectra do not possess exact $C_6$ rotational symmetry. Increasing the bond dimension produces no appreciable change in the locations of the dominant features or in the broad momentum dependence at any of the three energies. The features used in the main-text comparison are therefore converged at the resolution shown.

\section{Additional comparisons of Neutron Data against DMRG and DDMRG Results}

\subsection{Quantitative measure of agreement between neutron data and simulations}

To quantify the relative contributions of the two simulated QSLs, we construct a mixed structure factor
\begin{equation}
\mathcal{S}_{w}
=
w\mathcal{S}_{\mathrm{iso}}
+
(1-w)\mathcal{S}_{\mathrm{chev}},
\end{equation}
where $\mathcal{S}$ denotes either $S(\mathbf{Q})$ or $S(\mathbf{Q},\omega)$ and $w$ is the isotropic weight. Thus, $w=0$ corresponds to the pure chevron state and $w=1$ to the pure isotropic state. Each mixture was processed using the same momentum
mask, symmetrization, magnetic form factor, and intensity-scaling procedure as the corresponding main-text comparison.

\begin{figure*}[htbp]
\centering
\includegraphics[width=0.95\textwidth]{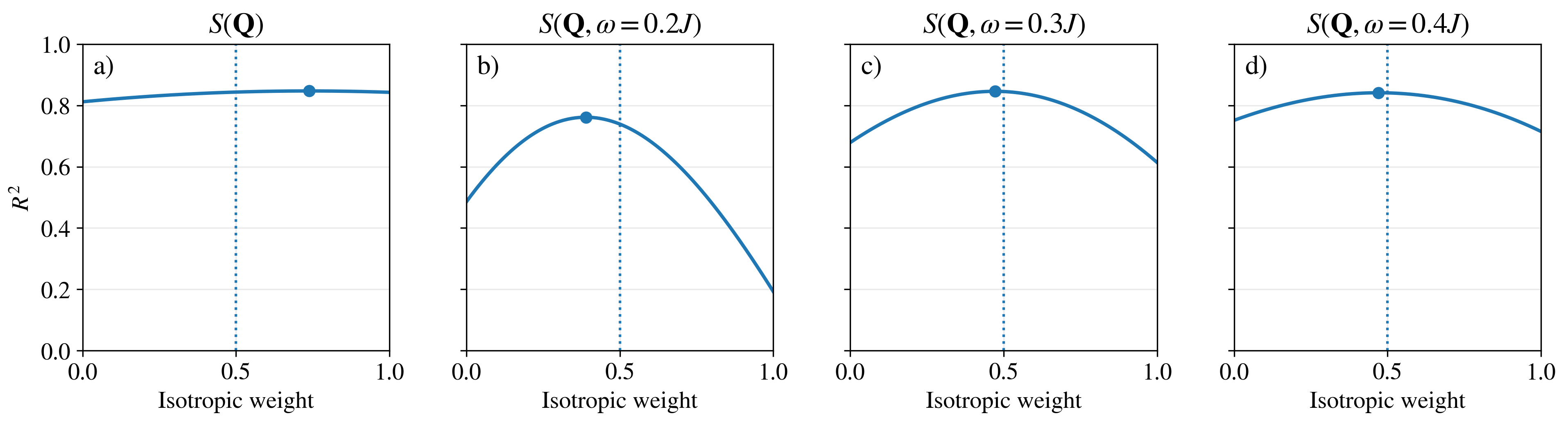}
\caption{\textbf{Dependence of the neutron--simulation agreement on
the isotropic weight.}
Coefficient of determination $R^2$ between the neutron data and the
mixed-state model as the isotropic weight $w$ is varied from the pure
chevron state ($w=0$) to the pure isotropic state ($w=1$).
\textbf{(a)} Equal-time structure factor $S(\mathbf{Q})$.
\textbf{(b--d)} Dynamical structure factors at
$\omega=0.2J_1$, $0.3J_1$, and $0.4J_1$, respectively.
Circular markers denote the maximum $R^2$ in each panel, and dotted
vertical lines mark the equal mixture, $w=0.5$.}
\label{sifig:si_rsq}
\end{figure*}

As shown in Fig.~\ref{sifig:si_rsq}(a), the agreement based on the equal-time structure factor varies only weakly with the isotropic
weight. Although the formal maximum occurs at a somewhat larger isotropic fraction, the small variation in $R^2$ means that $S(\mathbf{Q})$ alone does not tightly constrain the relative population of the two states. The energy-resolved structure factors are substantially more sensitive to the mixture. In particular, $S(\mathbf{Q},\omega=0.2J_1)$ gives a pronounced maximum near an equal admixture, while the maxima at $0.3J_1$ and $0.4J_1$ also remain close to equal weight. The dynamical response therefore provides the stronger constraint on the relative contributions of the isotropic and chevron correlations.

\subsection{Dynamical spin structure factor $S(\mathbf{Q},\omega)$ at high-symmetry points } 

\begin{figure*}[htbp]
\centering
\includegraphics[width=0.95\textwidth]{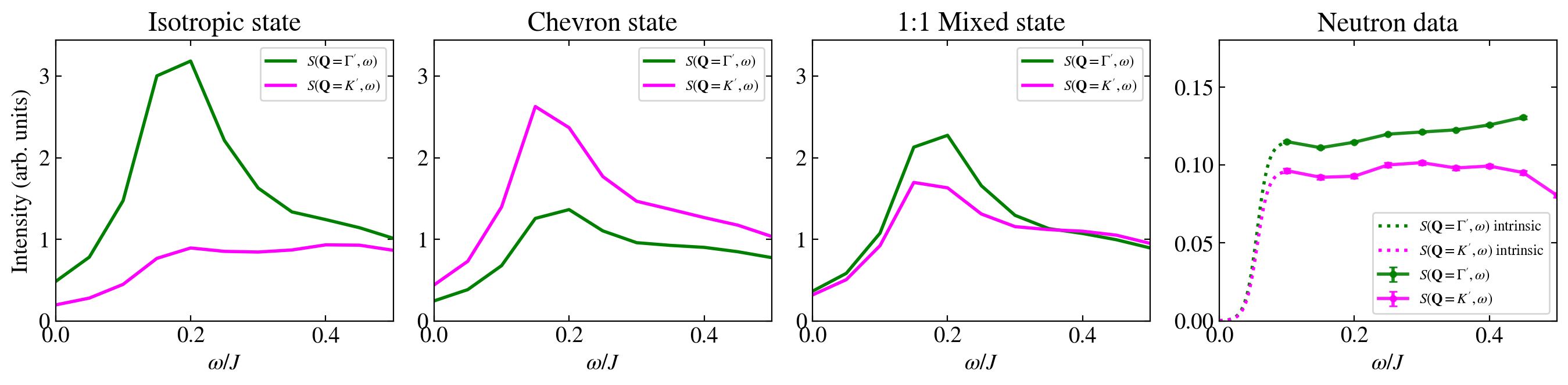}
\caption{\textbf{Energy dependence of the dynamical structure factor at $\Gamma'$ and $K'$.} \textbf{(a--c)} DDMRG $S(\mathbf{Q},\omega)$ on the XC12-12 cylinder at $\Gamma'$ and $K'$ for the isotropic QSL, chevron QSL, and their equal-weight average, respectively.
\textbf{(d)} Corresponding neutron scattering intensity at the same high-symmetry regions, with the experimental energy converted to $\omega/J_1$ using $J_1=15$~meV. Solid curves with markers show the full corrected neutron response, while dotted curves show the intrinsic kagome component obtained after empirical separation of the impurity contribution. The simulation and neutron panels use separate vertical intensity scales.}
\label{sifig:sqw_highsym}
\end{figure*}

Figure~\ref{sifig:sqw_highsym} compares the energy dependence of the DDMRG intensity at $\mathbf{Q}=\Gamma'$ and $\mathbf{Q}=K'$ on the XC8-12 cylinder, with resolution of $0.05J_1$. In the isotropic QSL, the $\Gamma'$ intensity shows a pronounced peak at $0.15J_1$, slightly larger than the $\sim0.1J_1$ triplet gap obtained from DMRG, whereas the $K'$ intensity is weaker at low energy and has no comparable peak. The situation is reversed in the chevron QSL, for which the low-energy intensity is concentrated at $K'$.

An equal-weight average reduces this strong momentum contrast and produces comparable spectral weight and broadly similar energy dependence at $\Gamma'$ and $K'$. This behavior resembles the neutron data, in which the intrinsic responses at the two momenta are also similar. The experimental response does not, however, exhibit a comparably sharp peak immediately above the triplet-gap scale and instead remains broad over the measured energy range. Because the simulation and experimental panels have different vertical scales and different energy broadenings, this comparison concerns the relative momentum and energy dependence rather than the absolute intensities.

\subsection{Dynamical spin structure factor $S(\mathbf{Q},\omega=0.2J)$ for XC12-12 cylinder}

\begin{figure*}[htbp]
\centering
\includegraphics[width=0.8\textwidth]{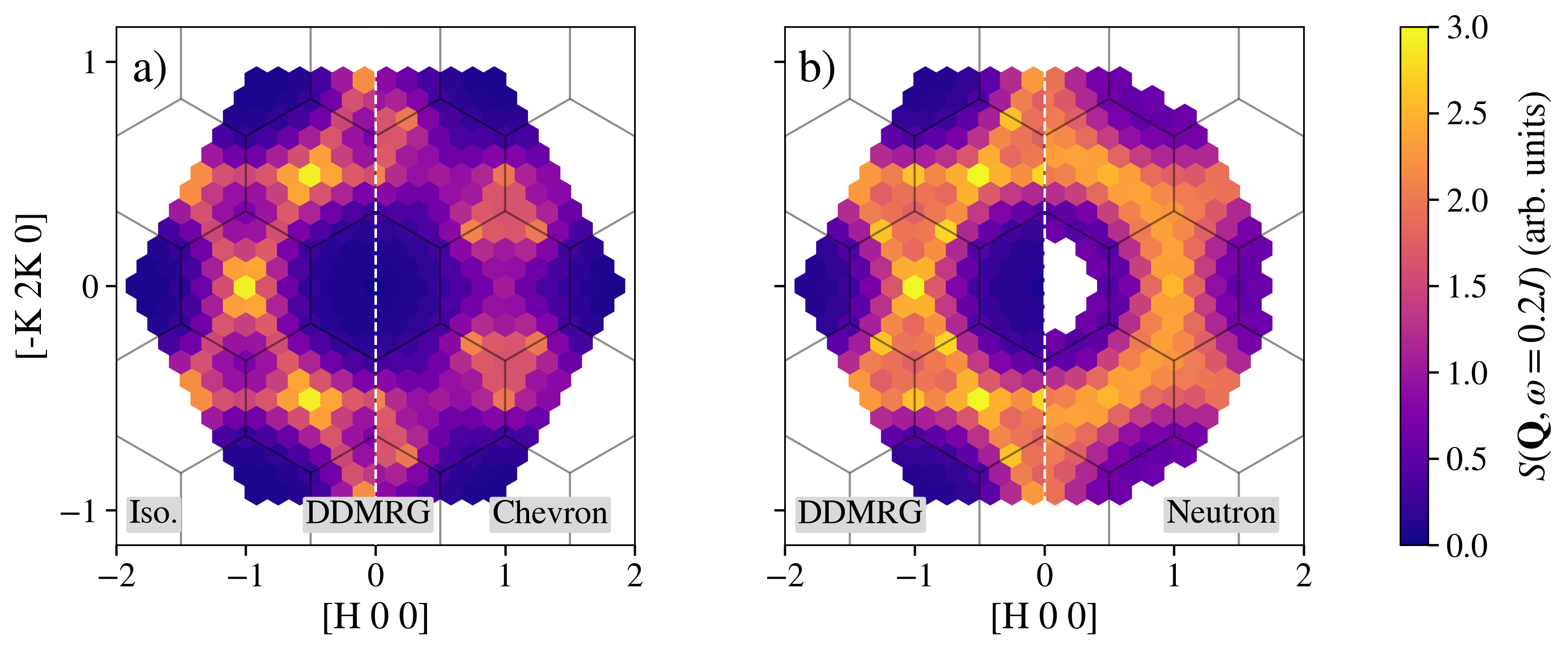}
\caption{\textbf{XC12-12 DDMRG comparison at
$\omega=0.2J_1$.} \textbf{(a)} $D_6$-symmetrized dynamical structure factors for the isotropic QSL (left half) and chevron QSL (right half), calculated on the XC12-12 cylinder with maximum bond dimension $m=9000$. \textbf{(b)} Equal-weight average of the two simulated structure factors (left half) compared with the neutron scattering data (right half). The Cu$^{2+}$ magnetic form factor and the same intensity-scaling convention used in the main-text comparison are applied to the simulated spectra.}
\label{sifig:xc12_dssf}
\end{figure*}

To test whether the simulated $S(\mathbf{Q},\omega)$ depends on the cylinder width, we also calculated it on an XC12-12 cylinder and kept a maximum bond dimension of $m=9000$. Because the simulation is significantly more challenging on an XC12 cylinder, we picked one representative $\omega=0.2$, and the results are shown in Fig.~\ref{sifig:xc12_dssf}. The qualitative features remain the same: the isotropic QSL has enhanced intensity at $\Gamma'$, while the chevron QSL has spectral weight concentrated near $K'$. Their equal-weight average suppresses both sets of peaks and produces a diffuse structure that agrees substantially better with the neutron data than either state alone. The two-state interpretation is therefore generic and not specific to XC8 cylinder.

\bibliography{refs}